\documentclass[journal]{IEEEtran}
\usepackage{amsthm}
\usepackage[utf8]{inputenc}
\usepackage[english]{babel}
\usepackage{amsmath, amsfonts, amssymb}
\usepackage{soul}

\usepackage{array}
\usepackage{multirow}

\newcolumntype{C}[1]{>{\centering\arraybackslash}p{#1}}
\newcolumntype{L}[1]{p{#1}}

\DeclareFontFamily{U}{BOONDOX-calo}{\skewchar\font=45 }
\DeclareFontShape{U}{BOONDOX-calo}{m}{n}{
  <-> s*[1.05] BOONDOX-r-calo}{}
\DeclareFontShape{U}{BOONDOX-calo}{b}{n}{
  <-> s*[1.05] BOONDOX-b-calo}{}
\DeclareMathAlphabet{\mathcalboondox}{U}{BOONDOX-calo}{m}{n}
\SetMathAlphabet{\mathcalboondox}{bold}{U}{BOONDOX-calo}{b}{n}
\DeclareMathAlphabet{\mathbcalboondox}{U}{BOONDOX-calo}{b}{n}

\DeclareFontFamily{OT1}{pzc}{}
\DeclareFontShape{OT1}{pzc}{m}{it}{<-> s * [1.000] pzcmi7t}{}
\DeclareMathAlphabet{\mathpzc}{OT1}{pzc}{m}{it}

\usepackage{multidef}
\usepackage{mathtools}

\multidef[prefix=set]{\ensuremath{\mathbb{#1}}}{A-Z}

\multidef[prefix=cal]{\ensuremath{\mathcal{#1}}}{A-Z}

\multidef[prefix=v]{\ensuremath{\mathbf{#1}}}{a-z}

\multidef[prefix=m]{\ensuremath{\mathbf{#1}}}{A-Z}
\newcommand{\mLambda}{\boldsymbol{\Lambda}}

\multidef[prefix=layer]{\ensuremath{\mathpzc{#1}}}{a-z}

\newcommand{\ones}{\boldsymbol{1}}
\newcommand{\zeros}{\boldsymbol{0}}

\newcommand{\diag}{\mathrm{diag}}
\newcommand{\trace}{\mathrm{tr}}

\usepackage{booktabs}
\usepackage{tabularx}
\newcolumntype{Y}{>{\centering\arraybackslash}X}
\usepackage{xcolor}
\usepackage[colorlinks=true, urlcolor=blue, linkcolor=red]{hyperref}
\usepackage{subcaption}
\usepackage{algorithm,algorithmic,gensymb}

\newcommand*{\matr}[1]{\mathbf{#1}}
\theoremstyle{assumption}
\newtheorem{assumption}{Assumption}
\begin{document}

\title{Learning Graph Filters for Structure-Function Coupling based Hub Node Identification}

\author{ Meiby Ortiz-Bouza$^{\dagger}$, Duc Vu$^{\dagger}$, Abdullah Karaaslanli, and Selin Aviyente,~\IEEEmembership{Senior Member,~IEEE}
        
\thanks{This work was supported in part by the National Science Foundation under Grants CCF-2211645 and CCF-2312546.}
\thanks{$\dagger$ The first two authors contributed equally to this manuscript.}
\thanks{The authors are with the Department of Electrical and Computer Engineering, Michigan State University, East Lansing, MI 48824 USA (e-mail: ortizbou@msu.edu; vuduc2@msu.edu; karaasl1@msu.edu; aviyente@egr.msu.edu).}}

\markboth{}%
{Shell \MakeLowercase{\textit{et al.}}: A Sample Article Using IEEEtran.cls for IEEE Journals}


\maketitle
\begin{abstract}
Over the past two decades, tools from network science have been leveraged to characterize the organization of both structural and functional networks of the brain. One such measure of network organization is hub node identification.
Hubs are specialized nodes within a network that link distinct brain units corresponding to specialized functional processes. Conventional methods for identifying hub nodes utilize different types of centrality measures and participation coefficient to profile various aspects of nodal importance. These methods solely rely on the functional connectivity networks constructed from functional magnetic resonance imaging (fMRI), ignoring the structure-function coupling in the brain. 
In this paper, we introduce a graph signal processing (GSP) based hub detection framework that utilizes both the structural connectivity and the functional activation to identify hub nodes. The proposed framework models functional activity as graph signals on the structural connectivity. Hub nodes are then detected based on the premise that hub nodes are sparse, have higher level of activity compared to their neighbors, and the non-hub nodes' activity can be modeled as the output of a graph-based filter. Based on these assumptions, an optimization framework, GraFHub, is formulated to learn the coefficients of the optimal polynomial graph filter and detect the hub nodes. The proposed framework is evaluated on both simulated data and resting state fMRI (rs-fMRI) data from Human Connectome Project (HCP).
\end{abstract}
\section{Introduction}
\label{sec:intro}
The recent developments in the field of human connectome research \cite{bullmore2009complex, sporns2007identification} provide us the opportunity to unravel the topological characteristics of brain networks using graph-theoretic approaches. Both the structural and functional brain systems can be characterized using tools from complex network theory such as 
small-world topology, highly connected hubs and modularity \cite{bassett2017small,sporns2016modular}. In this line of research, the brain is modeled as a graph composed of nodes and edges. The nodes represent neurons or brain regions and the edges represent physical connections or statistical associations between regions \cite{biswal1995functional} for structural and functional networks, respectively. Graph-theoretic methods offer measures to depict the features of the network, including modules \cite{he2009uncovering,yeo2011organization} and hubs \cite{buckner2009cortical,power2013evidence}. 

Hubs are defined as densely connected regions in human brain networks and play a crucial role in global brain communication \cite{van2013network} and support a broad range of cognitive tasks, such as working memory \cite{liu2017intrinsic} and semantic processing \cite{xu2016intrinsic}. Growing evidence suggests that these highly connected brain hubs are preferentially targeted by many neuropsychiatric disorders \cite{dai2015identifying}, providing critical clues for understanding the biological mechanisms of disorders and establishing biomarkers for disease diagnosis and treatment \cite{fornito2015connectomics}. 

Traditionally, hubs have been defined as nodes with high degree or high centrality based on functional connectivity networks. Node degree is the simplest and most commonly used means of identifying hubs in graphs. However, it has been shown that this approach is problematic in correlation based networks such as the functional connectivity networks \cite{power2013evidence}. The influence of community size on
degree and the susceptibility of degree to distortion in volume-based brain networks result in biased estimates of hub nodes. For this reason, other centrality metrics based on the combination
of degree and path length, e.g., betweenness, closeness, eigenvector, and PageRank centralities, have been proposed to characterize hubs \cite{joyce2010new}. Others have used the node role approach that relies on the community structure of the functional network. Namely, centrality measures
identify hubs and participation coefficients using within-module degree z-score then
classify hub type, e.g., provincial vs. connector hubs \cite{he2009uncovering,pedersen2020reducing}. However, all of these methods rely only on the functional connectivity network without considering the coupling between the brain's anatomical wiring, i.e., structural network, and its dynamic functional properties, e.g., the BOLD signal \cite{fotiadis2024structure}. 

Recently, tools from the field of graph signal processing (GSP) have been adapted to analyze functional brain signals with respect to the structural connectivity graphs they reside on  \cite{huang2018graph}.
Concepts of
graph Fourier transform (GFT) and the corresponding notions of graph spectral components and graph filtering have been utilized to analyze brain signals with respect to the spectrum of the underlying structural connectivity.
These GSP tools permit the decomposition of a graph signal into different frequency components that
represent different levels of variability 
 and have been used for quantifying structural-functional coupling, dimensionality reduction, and classification \cite{huang2018graph,preti2019decoupling,glomb2020connectome,menoret2017evaluating,glomb2021functional,margulies2016situating}. 

 Inspired by these advances, in this paper we introduce a GSP-based framework for hub node identification in brain networks utilizing both the structural connectome and functional BOLD signals. The proposed approach is based on learning the optimal graph filter for detecting hub nodes with the following assumptions: (i) hub nodes are sparse and have high activation patterns simultaneously with a more diverse set of connections, i.e., their activity corresponds to the high-frequency component of the BOLD signal, and (ii) the non-hub nodes' activation patterns are low-frequency/smooth with respect to the structural connectome, thus can be modeled as the output of a graph diffusion kernel, e.g., polynomial graph filter. These assumptions are incorporated into a general optimization framework where the smoothness and sparsity are quantified by graph total variation and $\ell_{1}$ norm, respectively. Once the optimal graph filter is learned, a hub scoring function based on the local gradient of the nodes is introduced to identify the hub nodes. Participation coefficient is used to further identify the connector hubs. The proposed method is evaluated on both simulated data and rs-fMRI data from HCP. The results are compared to the state-of-the-art hub node identification methods and recently published meta-analysis of hub nodes in rs-fMRI \cite{xu2022meta}.

 The main contributions of the proposed work are three-fold. First, unlike existing hub node detection methods that rely on only the connectivity graphs, the proposed method, GraFHub, incorporates the structural connectivity and the functional activation signals into the same framework thus taking the structure-function coupling in the brain into account \cite{fotiadis2024structure}. Second, in addition to introducing a GSP-based learning framework, this paper also employs two different metrics for hub node scoring, i.e., reconstruction error and smoothness, as well as two different methods for identifying the hubs, i.e., thresholding vs. rank ordering. While reconstruction error is commonly used in traditional anomaly detection, the smoothness metric quantifies the local gradient of the graph signal with respect to the graph and thus quantifies the hub score by taking both the graph signal and the graph structure into account. Finally, by learning the optimal graph filter for separating hub nodes from non-hub nodes, GraFHub provides interpretability to hub node identification. In particular, we show a strong correlation between the average hub score for a given brain network, the graph spectrum of the functional activation signals and the shape of the learned filters. 
 %
%
\section{Related Work}
\subsection{Hub node identification}
Current hub node identification methods can be grouped into two categories. The first category of methods determines hubs based on node centrality. These methods sequentially select a set of hub nodes by ranking a nodal centrality metric such as degree \cite{nijhuis2013topographic},
clustering coefficient \cite{onnela2005intensity}, vulnerability \cite{kaiser2004edge}, betweenness \cite{zalesky2010network}, and eigenvector centrality \cite{lohmann2010eigenvector}.
However, detecting hubs only using high nodal centrality ignores the interdependencies in the networks resulting in the detection of provincial
hubs, instead of connector hubs
which predominantly connect nodes across different modules. The second group of methods uses module-based methods that identify hub nodes based on the network modularity \cite{van2013network}. These methods detect hub nodes by first identifying the modular organization of the network using a community detection algorithm \cite{fortunato2010community}. Connector hub nodes are then detected based on the diversity of connections associated with the module partition. Although this method initially considers global network properties, the final hub detection uses a sorting-based method. Moreover, the optimality of the detected modules is not guaranteed \cite{fortunato2010community}. In recent years, alternatives to these two categories of methods have been proposed by combining multiple approaches, such as degree and participation coefficient, to obtain more reliable estimates of hubs \cite{jiao2018hub}. 

Recently, graph spectral methods have been proposed to detect the hubs such that the removal of the identified hubs results in a network with multiple connected components, or equivalently an increase in the number of $0$-eigenvalues in the graph Laplacian spectrum \cite{yang2019joint, yang2021joint}. GFT has also been employed to define a measure of centrality called GFT centrality (GFT-C) \cite{singh2017gft}. GFT-C first defines an importance signal for each node based on the shortest paths between that node and the other nodes. Hub scores are then determined by the weighted sum of GFT coefficients of the importance signal where the weight function is a pre-determined high-pass filter. Both graph spectral methods and GFT-C still rely on only the connectivity graph, i.e., the structural or functional connectome, without considering the coupling between the two modalities. 
%
\vspace{-0.1in}
\subsection{Graph Filter Learning}
Graph filtering offers an extension of conventional filtering approaches to signals defined in the non-Euclidean domain, e.g., irregular data structures arising in biological, financial, social, economic, sensor networks etc. \cite{dong2020graph, isufi2024graph}. Graph filters are information processing architectures tailored to graph-structure data and have been used for many signal processing tasks, such as denoising \cite{chen2014signal}, signal recovery \cite{marques2015sampling, tanaka2020sampling}, classification \cite{chen2014semi}, and anomaly
detection \cite{drayer2019detection}. The design of graph filters to obtain a desired
graph frequency response has been studied and analyzed in prior work \cite{liu2018filter,segarra2017optimal}. More recently, the problem of learning the optimal graph filter for a given task has been addressed. For example, in \cite{ramirez2021graph}, the problem of blind deconvolution is addressed where the observed signals are modeled as the output of a graph filter and both the filter and the input signal are learned simultaneously. In \cite{kroizer2022bayesian}, the problem of random graph signal estimation from a nonlinear observation model is addressed with an estimator that is parameterized in terms of shift-invariant graph filters. In all of these cases, the goal is graph signal recovery or reconstruction, minimizing the mean square error and not to identify the outlying nodes.

Closely related to the problem of hub node identification, spectral graph filtering has recently been employed in node-based anomaly detection. In particular, spectral properties of anomalies have been analyzed using graph signal processing concepts and graph spectral filters \cite{egilmez2014spectral,francisquini2022community,tang2022rethinking,gao2023addressing}. In \cite{egilmez2014spectral}, graph-based filters are employed to project graph signals onto normal and anomaly subspaces, and a thresholding mechanism is used to label anomalous instances. In \cite{francisquini2022community}, a community based anomaly detection approach is proposed using spectral graph filters. However, both of these methods use pre-determined filters such as the ideal low-pass/high-pass filters with no optimization of the filter shapes. More recently, the problem of estimating network centrality from the data observed on the nodes has been addressed \cite{he2020estimating,roddenberry2021blind}. The data supported on the nodes is modeled as the output of a graph
filter applied to white noise and centrality rank is learned without inferring the graph structure. This formulation reduces to determining centrality based on the principal components of the observed data's covariance matrix without taking the graph structure into account. Our proposed approach is similar to this line of work in the way it models the normal activity as the output of a graph filter. However, in our framework the graph structure is known and the normal activity is the output of an unknown graph filter where the input is observed. 



%

\section{Background}
\label{sec:background}
%
\subsection{Notations}
\label{ssec:notations}
In this work, scalars are represented with lowercase and uppercase letters, e.g., $n$ or $N$. Lowercase and uppercase bold letters, e.g., $\vx$ and $\mX$, are employed for vectors and matrices, respectively. $i$th entry of a vector $\vx$ is shown by $x_{i}$. For a matrix $\mX$, $X_{ij}$, $\mX_{i\cdot}$ and $\mX_{\cdot i}$ represent its $ij$th entry, $i$th row and $i$th column, respectively. All-ones vector, all-zeros vector and identity matrix are shown as $\ones$, $\zeros$, and $\mI$, respectively.  
$\trace(\cdot)$ and $^\top$ refer to the trace and the transpose of a matrix, respectively. 
For a vector $\vx \in \setR^{N}$, $\diag(\vx)$ is a diagonal matrix $\mX \in \setR^{N \times N}$ with $X_{ii} = x_i$. For a matrix $\mX \in \setR^{N \times N}$, $\diag(\mX)$ is a vector $\vx$ with $x_{i} = X_{ii}$. 
\vspace{-0.1in}
\subsection{Graphs and Graph Signals}
\label{ssec:graphs-and-signals}
An undirected weighted graph is represented as $G=(V, E, \mA)$ where $V$ is the node set with cardinality $|V| = N$ and $E$ is the edge set. $\mA \in \setR^{N\times N}$ is the adjacency matrix of $G$, where $A_{ij} = A_{ji}$ is the weight of the edge between nodes $i$ and $j$ and $A_{ij} = A_{ji} = 0$ if there is no edge between nodes $i$ and $j$. $\vd = \mA \ones$ is the degree vector and $\mD = \diag(\vd)$ is the diagonal degree matrix. The Laplacian matrix of $G$ is defined as $\mL = \mD - \mA$. The normalized Laplacian matrix $\matr{L}_{n}$ is defined as $\matr{L}_{n}=\matr{D}^{-1/2}(\matr{D}-\matr{A})\matr{D}^{-1/2}=\matr{I}-\matr{D}^{-1/2}\matr{A}\matr{D}^{-1/2}=\matr{I} - \matr{A}_{n}$, where $\matr{A}_{n}$ is the normalized adjacency matrix. Its eigendecomposition is $\mL_{n} = \mU \mLambda \mU^{\top}$, where columns of $\mU$ are eigenvectors and $\mLambda$ is the diagonal matrix of eigenvalues with $0 = \Lambda_{11} \leq \Lambda_{22} \leq \dots \leq \Lambda_{NN}$.

A graph signal defined on $G$ is a function $f: V \to \setR$ and can be represented as a vector $\vf \in \setR^{N}$ where $f_{i}$ is the signal value on node $i$. Akin to conventional signal processing, a graph signal can be studied using its Fourier domain representation, which can be derived using the graph shift operator (GSO). A GSO is an $N\times N$ dimensional matrix representing the structure of the graph, such as the adjacency, Laplacian or normalized Laplacian matrices \cite{shuman2013emerging}. In this work, the latter is employed as the GSO and its eigenvectors and eigenvalues are used to define GFT, where small eigenvalues correspond to low frequencies. Thus, GFT of $\vf$ is $\widehat{\vf} = \mU^\top \vf$ where $\hat{f}_{i}$ is the Fourier coefficient at the $i$th frequency component $\Lambda_{ii}$. When $P$ graph signals, $\{\vf_p\}_{p=1}^P$, are observed, they can be represented as a data matrix $\mF \in \setR^{N\times P}$ whose $p$th column $\mF_{\cdot p} = \vf_p$. 
In this case, GFT is represented as $\widehat{\matr{F}}=\mU^{\top}\matr{F}$.

GFT can be utilized to define a notion of signal variability over the graph such that $\mF$ is a smooth graph signal, i.e., $\mF$ has low variation over the graph if most of the energy of $\widehat{\mF}$ lies in the low-frequency components. The smoothness of $\mF$ can then be calculated using the total variation 
of $\mF$ measured in terms of the spectral density of its Fourier transform as:
\begin{align}
    \label{eq:smoothness}
    \trace(\widehat{\mF}^\top \mLambda \widehat{\mF}) = \trace(\mF^\top \mU \mLambda \mU^T \mF) = \trace(\mF^\top \mL_{n} \mF).
\end{align}
The quadratic term $\trace(\mF^\top \mL_n \mF)$ on the right-hand side of \eqref{eq:smoothness} is equal to $\sum_{i\neq j}A_{ij}(\frac{\mF_{i\cdot}}{\sqrt{d_i}}-\frac{\mF_{j\cdot}}{\sqrt{d_j}})^{2}$, whose smaller values indicate that the graph signal is smooth. In particular, for a smooth signal $\mF$, signal values on strongly connected nodes, i.e., large $A_{ij}$, are similar to each other.
%
%
\vspace{-0.05in}
\subsection{Graph Filtering}
\label{ssec:graph-filters}
A (convolutional) graph filter is an information processing unit that preserves specific properties of its input graph signal $\vf$ by applying a shift-and-sum operation on $\vf$ \cite{isufi2024graph}. Shifting $\vf$ requires propagating information in $\vf$ across the graph topology, which can be done by a linear transformation of $\vf$ with the GSO, e.g., $\mL_n\vf$, when the GSO is the normalized Laplacian. 
A graph filter of order $T$ is then defined as:
\begin{align}
    \calH(\mL_n) = \sum_{t=0}^{T-1} h_t \mL_n^t,
\end{align}
where $\matr{h} = [h_{0}, \dots, h_{T-1}]^{\top}$ is the vector of filter coefficients and $\mL_n^t$ is the $t$th power of the normalized Laplacian representing $t$-times shift operation. In the graph Fourier domain, the filtering operation can be interpreted as preserving the spectral content that is relevant to the task at hand. Namely, for a graph signal $\vf$, let $\tilde{\vf} = \mathcal{H}(\mL_n) \vf$. In graph Fourier domain:
\begin{align}
   \tilde{\vf}= \mU \mathcal{H}(\mLambda) \mU^\top \vf = \mU \mathcal{H}(\mLambda) \widehat{\vf} = \mU \widehat{\vf}_o,
\end{align}
where $\widehat{\vf}_o$ is the GFT of $\vf$ after being filtered by $\mathcal{H}(\mLambda)$, and whose $i$th Fourier coefficient is: 
\begin{align}
    [\hat{\vf}_o]_i = \mathcal{H}(\Lambda_{ii})\hat{f}_i = \sum_{t=0}^{T-1} h_t \Lambda_{ii}^t \hat{f}_i.
\end{align}   
Depending on the values of $\vh$, one can attenuate or amplify specific spectral components, thus, yielding an output graph signal $\tilde{\vf}$ that has a desired Fourier representation. For example, a smooth $\tilde{\vf}$ can be obtained by filtering out high-frequency components while amplifying low-frequency ones. This concept of graph filtering can be extended to multi-dimensional graph signals, $\mF \in \setR^{N\times P}$, such that the filtered signal is obtained as $\tilde{\mF}=\mU \mathcal{H}(\mLambda) \mU^\top \mF$.

%
\section{Optimal Graph Filtering for Hub Node Identification}
\label{sec:method}
\subsection{Problem Formulation}
\label{ssec:problem-formulation}
Given a graph with adjacency matrix $\matr{A}\in\mathbb{R}^{N\times N}$, and graph signal $\matr{F}\in\mathbb{R}^{N\times P}$, in this paper we focus on the detection of hub nodes by making the following assumptions.
\begin{assumption}
The hub nodes' activity is assumed to be high-frequency with respect to the graph.
\label{ass:smooth}
\end{assumption}
This assumption is inspired by the hypothesis that  ``hub regions
possess the highest level of activity" \cite{de2012activity}. This implies that hub nodes have different levels of activity compared to their neighbors. This existence of differences between neighboring nodes leads to the ‘right-shift’ of
spectral energy, which means the spectral energy distribution concentrates less in low frequencies and more in
high frequencies \cite{tang2022rethinking,chai2022can}.
\begin{assumption}
The normal nodes' activity can be modeled as the output of a graph based low-pass filter defined in terms of the GSO, $\matr{L}_{n}$.
\label{ass:diff}
\end{assumption}
This assumption is justified by work in GSP literature that models graph signals as diffusion processes, i.e., outputs of graph filters \cite{egilmez2018graph,segarra2017network}, and the fact that the non-hub nodes' activity can be modeled as low frequency on graphs (Assumption \ref{ass:smooth}).
\begin{assumption}
The number of hub nodes is much smaller than the normal nodes, i.e., hub nodes are sparse.
\label{ass:sparse}
\end{assumption}

Based on these assumptions, we model the graph signal $\matr{F}$ as $\tilde{\mF}+\matr{F}_{h}$, where $\tilde{\mF}$ corresponds to the normal (low-frequency) part and $\matr{F}_{h}$ corresponds to the hub (high-frequency) activity. Using Assumption \ref{ass:diff}, the normal part of the graph signal can be learned by filtering out high-frequency components in observed signals $\mF$, i.e., $\tilde{\matr{F}}=\mathcal{H}(\matr{L}_{n})\matr{F}$, where $\mathcal{H}(\matr{L}_{n})=\sum_{t=0}^{T-1} h_t\matr{L}^t_{n}$ is a polynomial graph filter to be learned. Since $\tilde{\mF}$ is considered to be smooth in Assumption \ref{ass:smooth}, the coefficients of $\mathcal{H}(\matr{L}_{n})$ are learned such that the total variation of $\tilde{\matr{F}}$ as calculated by \eqref{eq:smoothness} is minimized. Finally, using Assumption \ref{ass:sparse}, hub node activity, i.e., $\mF_h = \matr{F}-\tilde{\matr{F}}$, needs to be sparse.
The goal is to learn coefficients of $\calH(\mL_{n})$  such that all of these assumptions are satisfied simultaneously. This goal is formulated through the following optimization problem:
\begin{align}
\begin{split}
\label{eq:probform}
\underset{\matr{h},\matr{h}^\top \matr{h}=1}{\min} &\ \alpha ||\matr{F}-\tilde{\matr{F}}||_1+\text{tr}(\tilde{\matr{F}}^{\top}\matr{L}_{n}\tilde{\matr{F}}), \\
\text{s.t.\quad } &\ \tilde{\matr{F}}=\mathcal{H}(\matr{L}_{n})\matr{F}=\sum_{t=0}^{T-1} h_t\matr{L}_{n}^t\matr{F},
\end{split}
\end{align}
\noindent where $\matr{h}=[h_0, h_1,\cdots, h_{T-1}]$ is the vector of filter coefficients with the added constraint that $\matr{h}^\top\matr{h}=1$, i.e., the filter coefficients are normalized. The first term enforces the sparsity of the hub nodes (Assumption \ref{ass:sparse}), the second term quantifies the smoothness of the filtered signal (Assumption \ref{ass:smooth}), and $\alpha$ controls the trade-off between these two terms.

Expressing the filtered signal as $\tilde{\matr{F}}=\mathcal{H}(\matr{L}_{n})\matr{F}$, the problem in \eqref{eq:probform} becomes:
\begin{equation}
\label{eq:costfn}
\begin{split}
\underset{\matr{h},\matr{h}^\top \matr{h}=1}{\min} &\hspace{3mm}\alpha ||\matr{F}-\sum_{t=0}^{T-1} h_t\matr{L}_{n}^t\matr{F}||_1\\
&+\text{tr}\left(\matr{F}^\top\left(\sum_{t=0}^{T-1} h_t\matr{L}_{n}^t\right)\matr{L}_{n}\left(\sum_{t=0}^{T-1} h_t\matr{L}_{n}^t\right)\matr{F}\right).
\end{split}
\end{equation}
\vspace{-0.3in}
\subsection{Optimization}

In this section, we derive the solution to \eqref{eq:costfn} using ADMM \cite{parikh2014proximal}. 
By introducing an auxiliary variable $\matr{Z} =\matr{F}-\left(\sum_{t=0}^{T-1} h_t\matr{L}_{n}^t\right)\matr{F}$, the optimization problem is rewritten as
\begin{equation}
\begin{split}
\underset{\matr{h},\matr{Z}}{\min} \hspace{1mm}&\text{tr}\left(\matr{F}^\top\left(\sum_{t=0}^{T-1} h_t\matr{L}_{n}^t\right)\matr{L}_{n}\left(\sum_{t=0}^{T-1} h_t\matr{L}_{n}^t\right)\matr{F}\right)\\
+\alpha &||\matr{Z}||_1,\text{  s.t   } \matr{h}^\top \matr{h}=1, \matr{Z}=\matr{F}-\left(\sum_{t=0}^{T-1} h_t\matr{L}_{n}^t\right)\matr{F}.
\end{split}
\label{eq:ADMMopt}
\end{equation}

The corresponding scaled augmented Lagrangian is
\begin{equation}
\begin{split}
\label{eq:lag}
\mathcal{L}(\matr{Z},\matr{h},\matr{V})=\frac{\rho}{2}||\matr{Z}-(\matr{F}-\left(\sum_{t=0}^{T-1} h_t\matr{L}_{n}^t\right)\matr{F})+\matr{V}||_F^2\\
\text{tr}\left(\matr{F}^\top\left(\sum_{t=0}^{T-1} h_t\matr{L}_{n}^t\right)\matr{L}_{n}\left(\sum_{t=0}^{T-1} h_t\matr{L}_{n}^t\right)\matr{F}\right)
+\alpha ||\matr{Z}||_1,\\
\end{split}
\end{equation}
\noindent where $\matr{V}\in \mathbb{R}^{N\times P}$ is the Lagrange multiplier. The ADMM steps are then as follows.
\\ \\
\textbf{1.} $\matr{Z}$ update: The variable $\matr{Z}$ can be updated as
\begin{align}
\matr{Z}&^{k+1} = \operatorname*{argmin}_{\matr{Z}} \hspace{0.5mm} \mathcal{L}(\matr{Z},\matr{h}^{k},\matr{V}^{k}), \notag \\
=& \operatorname*{argmin}_{\matr{Z}} \hspace{0.5mm} \alpha ||\matr{Z}||_1+\frac{\rho}{2}||\matr{Z}-\matr{F}+\left(\sum_{t=0}^{T-1}h_{t}^{k}\matr{L}_{n}^{t}\right)\matr{F}+\matr{V}^k||_F^2, \notag \\ 
=& S_{\frac{\alpha}{\rho}}\left(\matr{F}-\left(\sum_{t=0}^{T-1} h_{t}^{k}\matr{L}_{n}^t\right)\matr{F}-\matr{V}^{k}\right),\hspace{0.52in}
\label{eq:Zupdate}
\end{align}
where $S_{\frac{\alpha}{\rho}}(\cdot)$ is the elementwise thresholding operator, which is the proximal operator of $\ell_1$ norm \cite{parikh2014proximal}.
\\ \\
\noindent \textbf{2.} $\matr{h}$ update: The filter coefficients $\matr{h}$ can be updated using:
\begin{equation*}
\begin{split}
\matr{h}^{k+1}&=\operatorname*{argmin}_{\matr{h},\matr{h}^\top \matr{h}=1}\mathcal{L}(\matr{Z}^{k+1},\matr{h},\matr{V}^{k}),\\
=&\operatorname*{argmin}_{\matr{h},\matr{h}^\top \matr{h}=1}\frac{\rho}{2}||\matr{Z}^{k+1}-\matr{F}+\left(\sum_{t=0}^{T-1} h_t\matr{L}_{n}^t\right)\matr{F}+\matr{V}^{k}||_F^2
\end{split}
\end{equation*}
\begin{equation}
\label{eq:hupdate}
+\text{tr}\left(\matr{F}^\top\left(\sum_{t=0}^{T-1} h_t\matr{L}_{n}^t\right)\matr{L}_{n}\left(\sum_{t=0}^{T-1} h_t\matr{L}_{n}^t\right)\matr{F}\right).
\end{equation}
Following the definitions in \cite{segarra2017optimal}, we define the $t$ times shifted input signal as $\matr{S}^{(t)} := \matr{U}\Lambda^t\matr{U}^\top\matr{F}$. We also define $\matr{S}_{(i)} \in \setR^{T \times P}$ corresponding to node $i$ where each row is the $t$ times shifted input signal at the $i$-th node, i.e. $[\matr{S}_{(i)}]_{t\cdot} := [\matr{S}^{(t)}]_{i\cdot}$. Hence, the filtered graph signal at node $i$ is $\tilde{\matr{F}}_{i\cdot} = \sum_{t=0}^{T-1}h_t [\matr{S}^{(t)}]_{i\cdot}= \matr{h}^\top \matr{S}_{(i)}$ and we have $\tilde{F}_{ip}=\matr{h}^\top \matr{s}_i^p$ where $\vs_{i}^p = [\mS_{(i)}]_{\cdot p}$. The objective function in \eqref{eq:hupdate} can then be written element-wise:
\begin{align*}
\mathcal{L}(\matr{Z}^{k+1}, \matr{h}, \matr{V}^{k}) =& 
    \frac{\rho}{2} \sum_{p=1}^P \sum_{i=1}^N (Z_{ip}^{k+1} - F_{ip} + \matr{h}^\top \matr{s}_i^p + V_{ip}^{k})^2 \\ &+ 
    \sum_{p=1}^P \sum_{i,j=1}^N (\matr{h}^\top \matr{s}_i^p)L_{ij}(\matr{h}^\top \matr{s}_j^p).  
\end{align*}
Taking derivative of $\mathcal{L}(\matr{Z}^{k+1}, \matr{h}, \matr{V}^{k})$ with respect to $\vh$ and equating it to $0$ yields $\matr{h}^{k+1} = -\matr{Y}^{-1}\matr{b}$, where 
\begin{align*}
    \matr{b} &= \rho\sum_{p=1}^P\sum_{i=1}^N\matr{s}_i^p(Z_{ip}^{k+1}-F_{ip}+V_{ip}^{k}),
\end{align*}
\begin{align*}
    \matr{Y} &= \sum_{p=1}^P(2\sum_{i,j=1}^N\matr{s}_i^pL_{ij}\matr{s}_j^{p^\top} +\rho\sum_{i=1}^N\matr{s}_i^p\matr{s}_i^{p^\top}).
\end{align*}
Finally, in order to satisfy the constraint $\matr{h}^\top\matr{h}=1$, $\matr{h}^{k+1}$ is projected onto the set defined by $\vh^\top \vh = 1$.
\\ 

\noindent \textbf{3.} $\matr{V}$ update: The Lagrangian multiplier can be updated using:
\begin{equation}
\begin{split}
\label{eq:Vupdate}
&\matr{V}^{k+1}=\matr{V}^{k}+\rho(\matr{Z}^{k+1}-\matr{F}+\left(\sum_{t=0}^{T-1} h_t^{k+1}\matr{L}_{n}^t\right)\matr{F}).
\end{split}
\end{equation}
\noindent These three variables are updated until convergence as described in Algorithm \ref{alg:gf}. Since our problem is a non-smooth convex optimization over a non-convex manifold $\vh^\top\vh=1$, when applying ADMM to the proposed optimization problem, there are no formal global convergence guarantees. However, ADMM is known to perform well on non-convex problems, often converging to locally optimal solutions \cite{boyd2011distributed, wang2019global}. Recent works also empirically show the convergence of ADMM for non-smooth problems over non-convex manifolds \cite{kovnatsky2016madmm}. 
\begin{algorithm}[h!]
\caption{GraFHub}
\label{alg:gf}
\begin{algorithmic}[1]
\renewcommand{\algorithmicrequire}{\textbf{Input:}}
\REQUIRE Adjacency matrix $\matr{A}$, graph signal $\matr{F}$, parameters $\alpha, \rho$, and filter order  $T$.
\renewcommand{\algorithmicrequire}{\textbf{Output:}} 
\REQUIRE $\tilde{\matr{F}}$, graph filter $\mathcal{H}(\matr{\Lambda})$.
\STATE $\matr{L}_{n}\leftarrow  \matr{I}-\matr{A}_n$
\STATE $[\matr{U},\matr{\Lambda}]\leftarrow \text{EVD}(\matr{L}_n)$
\STATE $\matr{S}^{(t)} \leftarrow \matr{U}\Lambda^t\matr{U}^\top\matr{F}$, $t\in\{0,1,\cdots,T-1\}$
\STATE $[\matr{S}_{(i)}]_{t\cdot} := [\matr{S}^{(t)}]_{i\cdot}$, for each node $i\in V$
\STATE Initialize $\matr{h}=\text{rand}(T,1)$, $\matr{V}=\text{rand}(N,P)$
\WHILE{$||\vh^{(k+1)}-\vh^{(k)}||^2>10^{-3}$}
\STATE update $\matr{Z}^{(k+1)}$ according to Eq. \eqref{eq:Zupdate}
\STATE update $\matr{h}^{(k+1)}$ according to Eq. \eqref{eq:hupdate}
\STATE update $\matr{V}^{(k+1)}$ according to Eq. \eqref{eq:Vupdate}
\ENDWHILE
\STATE $\tilde{\matr{F}}\leftarrow\matr{U}\left(\sum_{t=0}^{T-1}h_t^{(k+1)}\matr{\Lambda}^t\matr{U}^\top\right)\matr{F}$ 
\end{algorithmic}
\end{algorithm}
\subsection{Hub Scoring}
\label{ssec:hubscore}
Once the filtered graph signal $\tilde{\matr{F}}$ is obtained, hubs are scored using two different metrics. The first metric is the reconstruction error of the node attributes: 
\begin{align*}
\text{scores}(i)=||\matr{F}_{i\cdot}-\tilde{\matr{F}}_{i\cdot}||^2.
\end{align*}
The second hub scoring metric is based on the graph signal's local smoothness \cite{smith2017locating} before and after filtering:
\begin{equation*}
\begin{split}
    \text{scores}(i)=E(i)-\tilde{E}(i),
\end{split}
\end{equation*}
where $E(i)=\sum_{j=1}^N A_{ij}||\matr{F}_{i\cdot}-\matr{F}_{j\cdot}||^2$ and $\tilde{E}(i)=\sum_{j=1}^N A_{ij}||\tilde{\matr{F}}_{i\cdot}-\tilde{\matr{F}}_{j\cdot}||^2$ are the node gradients at node $i$ for the original and filtered signals, respectively. This metric quantifies the difference in the similarity of a node's value to its neighbors before and after filtering. It is hypothesized that for hub nodes, this difference will be larger as the hub nodes' activity tends to be dissimilar to its neighbors (Assumption \ref{ass:smooth}).

For both metrics, once the scores for each node $i\in V$ are computed, hubs are detected in two different ways: \textit{(i)} thresholding and \textit{(ii)} top $K$-hubs. For thresholding, we use the z-score approach, i.e., nodes whose z-score is larger than 3 are denoted as hubs. For the top $K$-hub approach, we consider the top-$K$ nodes with the highest scores as hubs \cite{yang2021joint}. 
In our experiments, $K$ is chosen as the point where there is a significant drop in the hub scoring metric, similar to the elbow criterion \cite{nainggolan2019improved}, as detailed in Section \ref{sec:experiments}.


\section{Experiments on simulated data}
\label{sec:experiments}
\subsection{Benchmark Methods}
In this section, we evaluate the performance of our method, GraFHub, on  simulated data. 
We compare the accuracy of GraFHub to three groups of commonly used hub node detection methods. The first group of methods relies only on the graph topology. This group includes graph-theoretic centrality measures such as the degree, eigenvector, and closeness centrality \cite{van2013network} and Graph Fourier Transform Centrality (GFT-C) \cite{singh2017gft}, a GSP based method that uses the GFT coefficients of an importance signal derived from the shortest path with respect to a particular node. 
The second group of methods only relies on graph signals and does not consider graph connectivity. This class includes statistical methods, such as Local Outlier Factor (LOF) \cite{Auskalnis_Paulauskas_Baskys_2018}, which uses data density to find outliers based on the distance between neighbors, and clustering methods, such as Isolation Forest \cite{liu2008isolation}. These methods learn an anomaly region and classify the nodes based on whether the node resides within the region or not. Unlike our method, none of these methods utilizes both the graph topology and the graph signals.

Finally, we compare our learning based method to a fixed graph filtering based method, graph high-pass filtering (GHF) \cite{zhu2021interpreting}. GHF utilizes the connectivity information and the graph signal to detect the hub nodes. GHF solves the following optimization problem to obtain the graph signals, $\tilde{\matr{F}}$, corresponding to the normal activity:
\begin{equation}
\min_{\tilde{\matr{F}}} \left\| \left( {\matr{I} + \beta \matr{L}_n} \right)^{1/2} \left( \tilde{\matr{F}}- {\matr{F}} \right) \right\|_F^2 + \frac{\xi}{2} \text{tr} \left( \tilde{\matr{F}}^T \matr{L}_n \tilde{\matr{F}} \right),
\end{equation}
\noindent where $\beta$ is a parameter that controls the cutoff frequency of the high-pass filter, and $\xi$ is the regularization parameter. Similar to GraFHub, GHF learns $\tilde{\matr{F}}$ that is smooth with respect to the underlying graph. Unlike GraFHub, this method does not learn the filter shape and does not enforce sparsity on the hub nodes. Additionally, to further demonstrate the superiority of learning the graph filter, we also compare our method against directly solving the optimization problem in \eqref{eq:probform} for $\tilde{\mF}$ as:
\begin{equation}
\begin{split}
\label{eq:probformF}
\underset{\tilde{\matr{F}}}{\min} \hspace{3mm} \alpha ||\matr{F}-\tilde{\matr{F}}||_1+\text{tr}(\tilde{\matr{F}}^{\top}\matr{L}_{n}\tilde{\matr{F}}).
\end{split}
\end{equation}

\subsection{Simulated Data}
\begin{figure*}[t]
    \centering
    \includegraphics[width=0.9\linewidth]{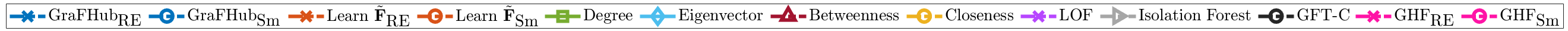} 
    \begin{minipage}{0.32\textwidth}
        \includegraphics[width=\linewidth]{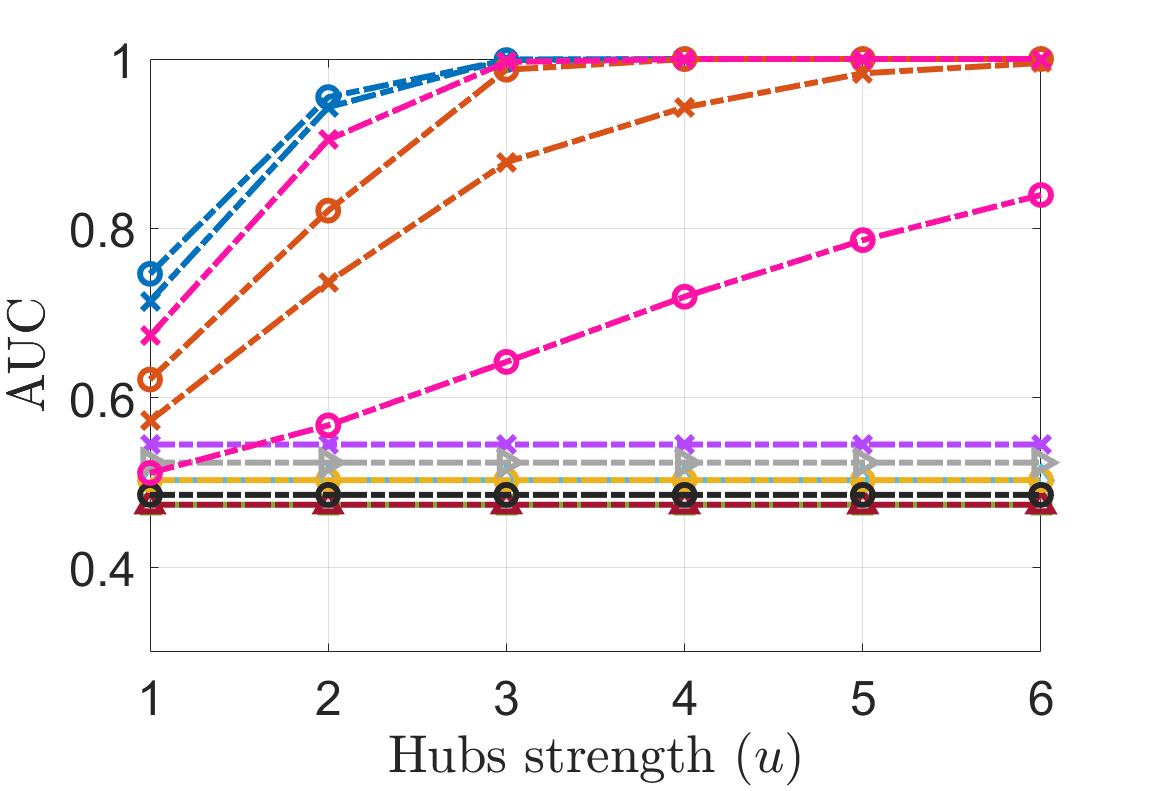}
         \caption*{(a)}
    \end{minipage}
    \hfill
    \begin{minipage}{0.32\textwidth}
        \includegraphics[width=\linewidth]{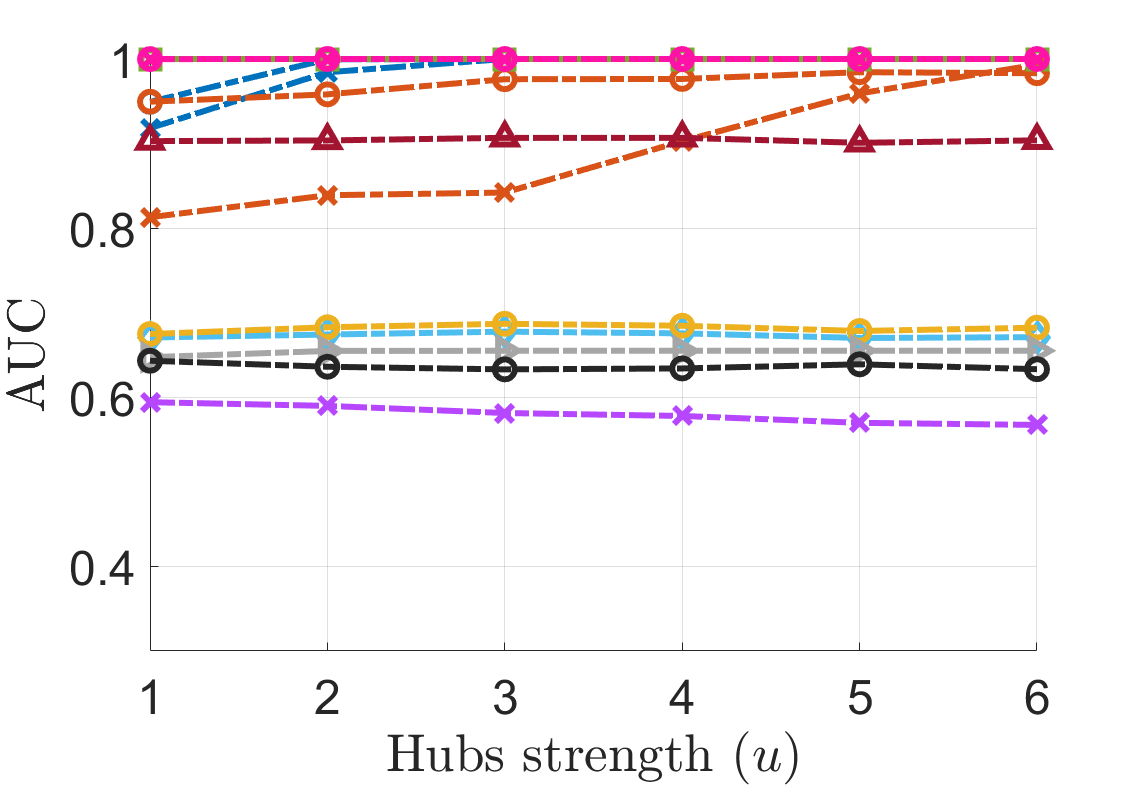} 
        \caption*{(b)}
    \end{minipage}
    \hfill
    \begin{minipage}{0.32\textwidth}
        \includegraphics[width=\linewidth]{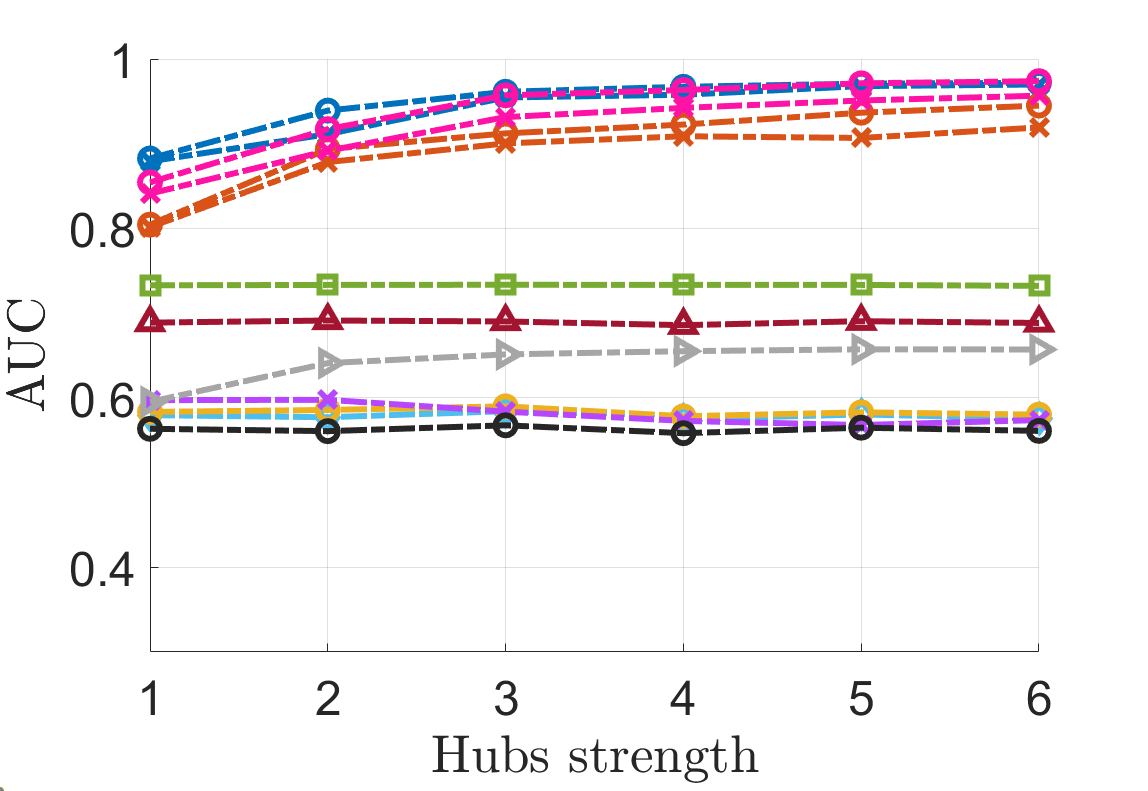}
        \caption*{(c)}
    \end{minipage}
    \begin{minipage}{0.32\textwidth}
        \includegraphics[width=\linewidth]{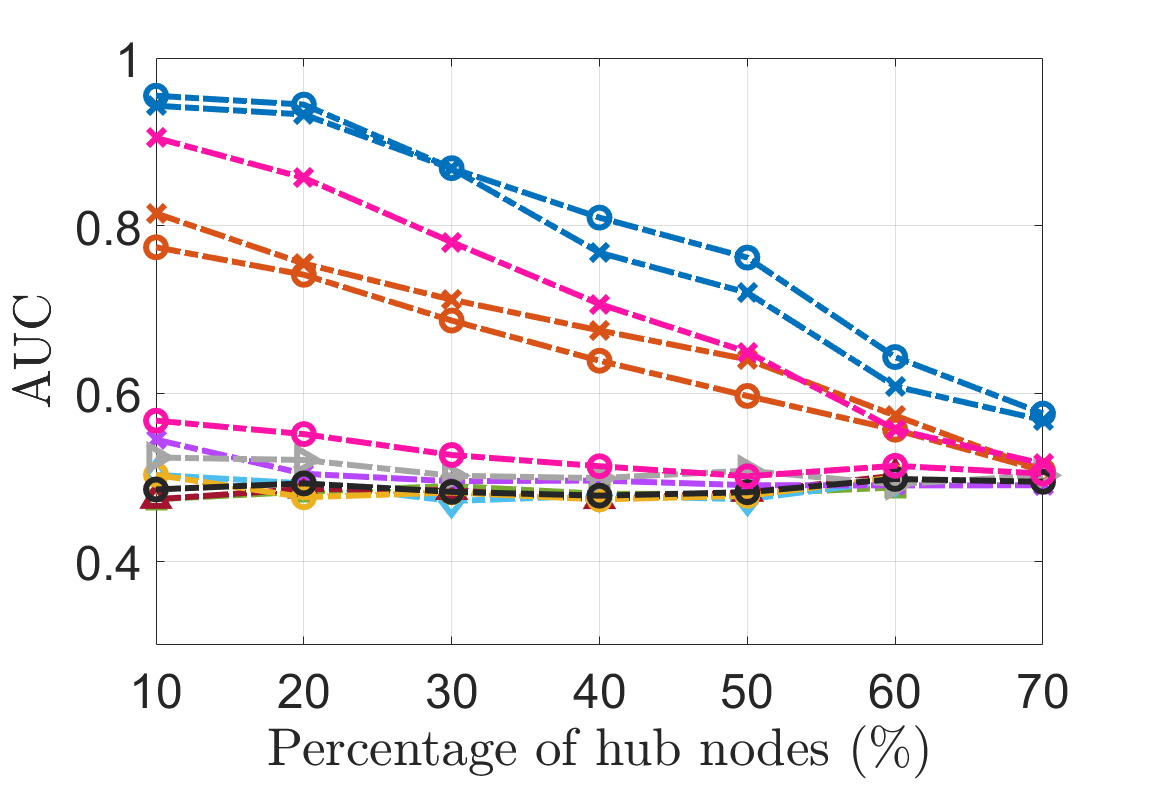}
        \caption*{(d)}
    \end{minipage}
    \hfill
    \begin{minipage}{0.32\textwidth}
        \includegraphics[width=\linewidth]{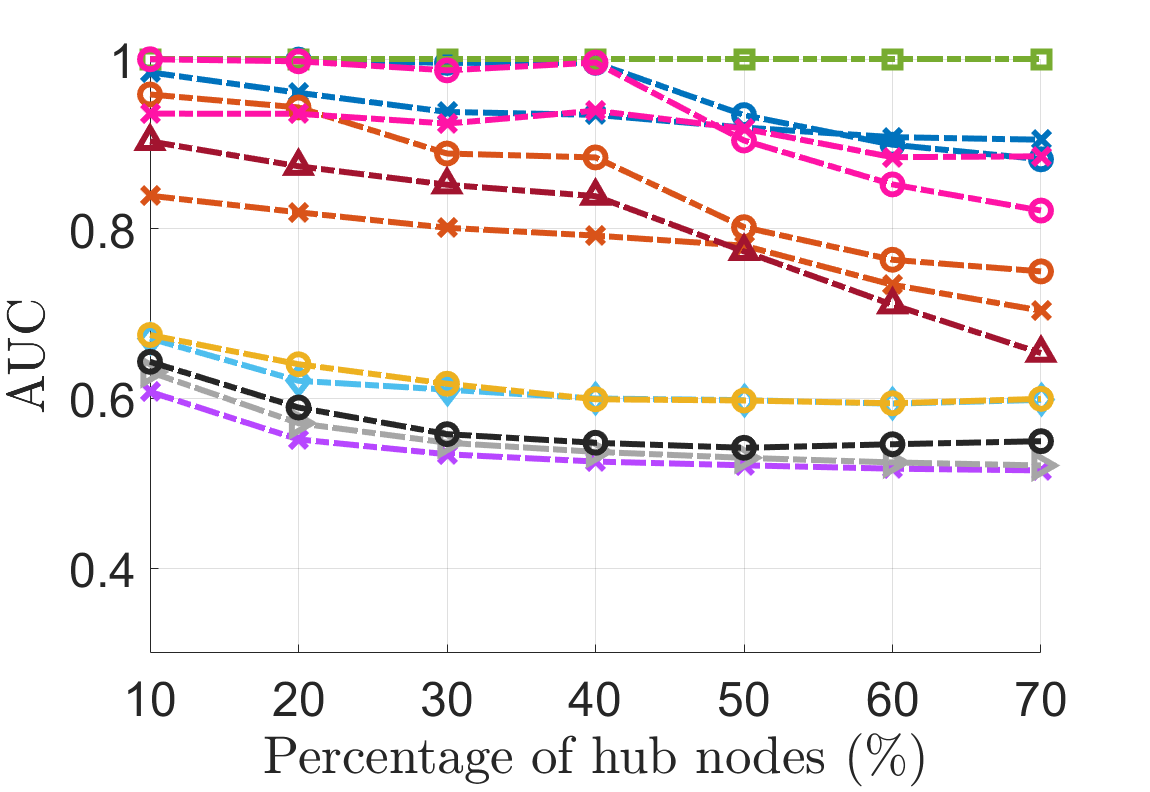} 
         \caption*{(e)}
    \end{minipage}
    \hfill
    \begin{minipage}{0.32\textwidth}
        \includegraphics[width=\linewidth]{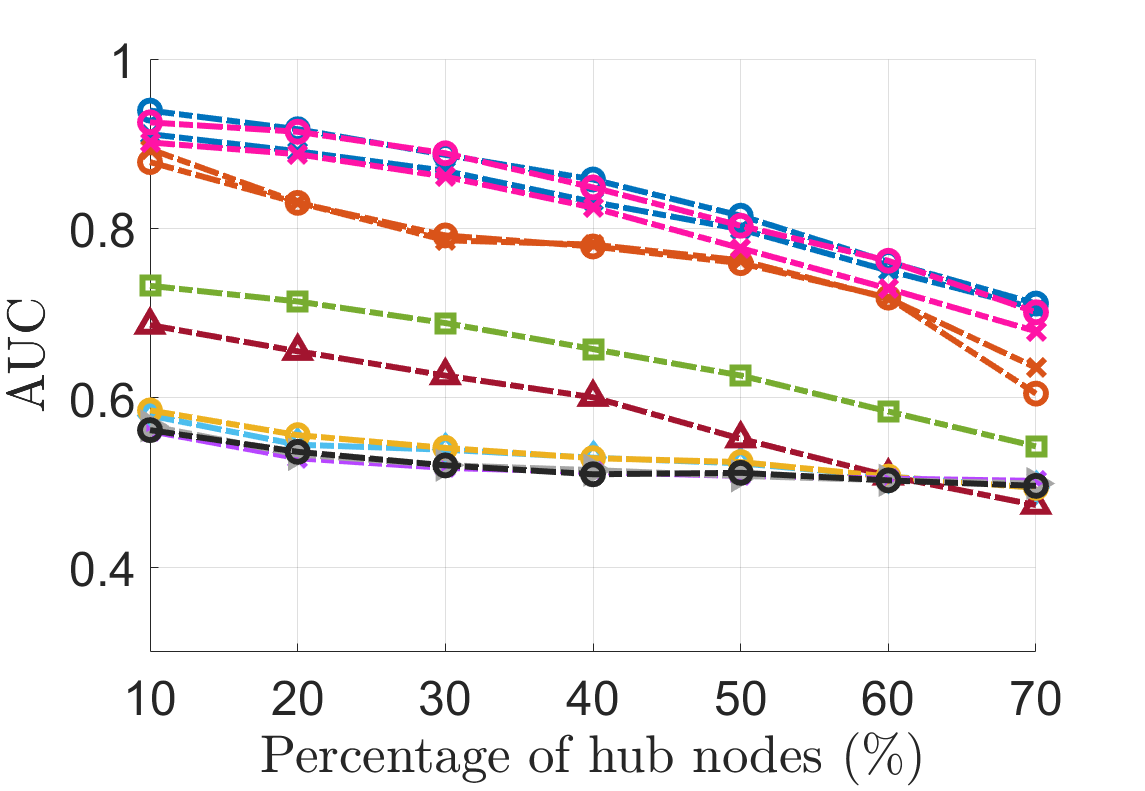}
        \caption*{(f)}
    \end{minipage}
    \caption{Performance of GraFHub on synthetic attributed graphs. AUC vs. Hub Signal Strength ($u$) (first row) and AUC vs. Percentage of nodes that are hubs (second row). From left to right, ER, BAdegree, and BAmixed models.}
    \label{fig:exp1}
\end{figure*} 

The simulated data are generated by first constructing a graph $G$ from either Erd\"{o}s-R\'{e}nyi (ER) or Barabási–Albert (BA) models. ER model creates a random graph where each edge is generated independently with probability $p$, resulting in a graph with no inherent structure, where edges are uniformly distributed. In our simulations, an ER random graph with $N$ nodes and edge density probability $p=0.1$ is generated. BA model follows a preferential attachment mechanism with growth parameter $m$, producing a scale-free graph. This means that the graph has a few highly connected nodes and many nodes with fewer connections, mimicking real-world networks like social and biological networks. In the following, a BA random graph with $N$ nodes and $m=3$ is generated. 

Once $G$ is constructed, $P$ smooth graph signals, $\matr{X}=[\matr{x}_1|\cdots|\matr{x}_P]$, are generated using Tikhonov filtering, i.e., $\matr{x}_{p}=(\gamma\matr{L}_n+\mI)^{-1}\matr{x_{0}}\in\mathbf{R}^{N}$, $p\in\{1,\ldots,P\}$, as the normal nodes' activity where $\matr{x_{0}}\sim \mathcal{N}(0,\matr{I})$ and $\gamma$ is the degree of smoothness. We then add synthetic hub nodes to $\mX$ by selecting a fixed percentage of the nodes as hubs. For the ER graph, these hub nodes are selected randomly, while for the BA graphs, they are selected in two different ways: \textit{(i)} hub nodes are selected as the nodes with the highest degree in the generated graph (BAdegree); \textit{(ii)} half of the hub nodes are selected randomly and the other half are selected from nodes with the highest degree (BAmixed). The signal values of selected hub nodes are perturbed by adding uniform noise in the interval $[-u\sigma, u\sigma]$ where $\sigma$ is the standard deviation of the norm of $\matr{X}_{\cdot i}$, and $u$ is the strength of the hubs. Unless noted otherwise, $N=1000$, $P=100$, $\gamma=30$, $u=2$, and the percentage of hub nodes is $10\%$. 

For all experiments, we report the performance of the aforementioned methods and two versions of the proposed method: $\textrm{GraFHub}_{\text{RE}}$ and $\textrm{GraFHub}_{\text{Sm}}$ employing reconstruction error and smoothness hub scores, respectively. The best $\alpha$ and $T$ values for GraFHub are determined from $\alpha\in[0.01, 0.02, 0.1, 0.2, 0.5, 1, 2, 5, 10, 20, 50, 100]$ and $T$ between 2 and 6. For GHF and the approach of learning $\tilde{\mathbf{F}}$ directly, the performance using both anomaly scoring metrics, RE and Sm, are reported, indicated by the subscripts RE and Sm.  The range of $\alpha$ is selected to be the same as GraFHub. For GHF, $\xi / 2 = 1/\alpha$, and $\beta=1$.  For GFT-C, the weights of the weighting function are computed as described in \cite{singh2017gft}. For signal-based anomaly detection methods, i.e., LOF and Isolation Forest, the default parameters are used. For LOF, the number of neighbors is set to 20. For Isolation Forest, number of estimators is set to 100. The results with the best performance are reported. The performance is quantified using Area Under the Receiver Operating Characteristic curve (AUC-ROC) where the hub scores returned by the methods are used without any thresholding or ranking. The average AUC-ROC over 50 runs is reported. 
\subsubsection*{Experiment 1: Strength of hubs}
In the first experiment, we vary the strength of the hubs, $u$. The top row of Figure \ref{fig:exp1} shows the results for ER (Figure \ref{fig:exp1} (a)), BAdegree (Figure \ref{fig:exp1} (b))and BAmixed (Figure \ref{fig:exp1} (c)) as $u$ is increased from $1$ to $6$. It can be seen that AUC-ROC score improves as $u$ increases for most of the methods that utilize graph signals for hub identification since the hubs become better separated from normal activity. Methods that rely only on network connectivity alone, such as centrality measures, show no change in performance as hub strength increases. These methods are inherently limited because they do not take the graph signal into account. An exception to this observation is the degree centrality for the BAdegree case, which has an AUC-ROC equal to 1 for all $u$ values as shown in Figure \ref{fig:exp1} (b). This is expected since hubs in BAdegree are defined by high connectivity, which aligns with the definition of hubs in degree centrality. However, its performance drops significantly for the ER and BAmixed models, where hub nodes are defined based on both graph signals and connectivity. This shows that degree centrality fails to detect hubs when the node’s importance is not purely based on connectivity. By combining structural connectivity and the graph signals, both $\textrm{GraFHub}_{\text{RE}}$ and $\textrm{GraFHub}_{\text{Sm}}$ achieve better AUC-ROC scores than the rest of the methods. GraFHub effectively captures the differences between hub and non-hub nodes by learning an optimal graph filter, allowing for better hub identification even when the hub nodes' strength is weak. While $\textrm{GraFHub}_{\text{RE}}$ and $\textrm{GraFHub}_{\text{Sm}}$ have similar performance in most cases, the latter is slightly better as it further eliminates false detection by using Assumption \ref{ass:smooth} in its scoring. Additionally, GraFHub outperforms the approach of learning $\tilde{\mF}$ directly from \eqref{eq:probformF} without learning a graph filter. This is due to the fact that our method constrains $\tilde{\mF}$ to be the output of a graph diffusion filter, thus it shapes its spectrum to be low-frequency. GHF performs similarly to our method for BAdegree; however, GraFHub outperforms GHF for ER and BAmixed cases as it learns the filter from the graph and the observed node activity unlike GHF which uses a pre-determined filter. 
\subsubsection*{Experiment 2: Percentage of hub nodes}
In the second experiment, we evaluate the performance of the methods as the percentage of hub nodes increases. The bottom row of Figure \ref{fig:exp1} shows AUC-ROC for ER (Figure \ref{fig:exp1} (d)), BAdegree (Figure \ref{fig:exp1} (e)), and BAmixed (Figure \ref{fig:exp1} (f)), where the percentage of hub nodes is increased from $10\%$ to $70\%$. The performance of most methods decreases as the percentage of hub nodes increases. In the case of GraFHub, this drop in performance can be attributed to one of the core assumptions of our method, i.e., the sparsity of hub nodes. The methods that rely only on connectivity information or node signal information are not affected by the number of hubs. However, they still perform worse than methods that use both connectivity and graph signals. Both $\textrm{GraFHub}_{\text{RE}}$ and $\textrm{GraFHub}_{\text{Sm}}$ have higher AUC-ROC scores than the rest of the methods for ER. For BAdegree and BAmixed, GraFHub and GHF show similar performance and they outperform the rest with the exception of degree centrality in the case of BAdegree, where hub definition aligns with degree centrality as discussed above. Similar to the previous experiment, $\textrm{GraFHub}_{\text{Sm}}$ has higher performance than $\textrm{GraFHub}_{\text{RE}}$ and learning a filter provides better hub detection than learning $\tilde{\mF}$ directly.   
\section{Application to Resting State fMRI Data}
\subsection{HCP Data}
The proposed method is applied on structural and functional neuroimaging data from 56 subjects collected as part of the Human Connectome Project (HCP)\footnote{\url{db.humanconnectome.org}}. 
Data acquisition is performed using a Siemens 3T Skyra with a 32-channel head coil \cite{van2013wu}. The scanning protocol includes high-resolution T1-weighted scans (256 slices, $0.7 \, \text{mm}^{3}$ isotropic resolution, $\text{TE} = 2.14 \, \text{ms}$, $\text{TR} = 2400 \, \text{ms}$, $\text{TI} = 1000 \, \text{ms}$, flip angle $=8^\circ$, $\text{FOV} = 224 \times 224 \, \text{mm}^2$, $\text{BW} = 210 \, \text{Hz/px}$, $\text{iPAT} = 2$) \cite{glasser2013minimal}. Diffusion data is collected with Spin-echo EPI, $\text{TR} = 5520 \, \text{ms}$, $\text{TE} = 89.5 \, \text{ms}$, flip angle $= 78^\circ$, refocusing flip angle $=160^\circ$, $\text{FOV} = 210 \times 180 \, \text{mm}^2 \, (\text{RO} \times \text{PE})$, matrix $= 168 \times 144 \, (\text{RO} \times \text{PE})$, 111 slices with thickness of $1.25 \, \text{mm}$ and $1.25 \, \text{mm}$ isotropic voxel size, multiband factor $3$, echo spacing $=0.78 \, \text{ms}$, $\text{BW} = 1488 \, \text{Hz/px}$, phase partial Fourier $=\frac{6}{8}$, b-values $=1000, 2000$, and $3000 \, \text{s/mm}^2$. Functional scans were collected using a multi-band sequence with MB factor $8$, isotropic $2 \, \text{mm}^3$ voxels, $\text{TE} = 33 \, \text{ms}$, $\text{TR} = 720 \, \text{ms}$, flip angle $= 52^\circ$, $\text{FOV} = 208 \times 180 \, \text{mm}^2 \, (\text{RO} \times \text{PE})$, 72 slices, $2.0 \, \text{mm}$ isotropic resolution, $\text{BW} = 290 \, \text{Hz/px}$, echo spacing $= 0.58 \, \text{ms}$ \cite{glasser2013minimal}. One hour of resting state data was acquired per subject in 15-minute intervals over two separate sessions with eyes open and fixation on a crosshair. Within each session, oblique axial acquisitions alternated between phase encoding in a right-to-left (RL) direction in one run and left-to-right (LR) in the other run. Minimally preprocessed images from HCP are used in the following analysis.
\vspace{-0.1in}
\subsection{Preprocessing}
Diffusion-weighted scans are analyzed using MRtrix3\footnote{\url{https://www.mrtrix.org/}} to construct the structural connectomes. The following operations are employed: multi-shell multi-tissue response function estimation, Glasser's multimodal cortical atlas parcellation, constrained spherical deconvolution, and tractogram generation with $10^6$ output streamlines. The volume is split into $N=360$ regions in two hemispheres (180 areas on the right and 180 areas on the left). In each hemisphere, Glasser divides 180 “areas” into 22 separate “regions”, which are referred to as the 22 larger partition cortices. Each of the 180 regions occupies one of 22 cortices which are displayed in a separate atlas. The number of fibers connecting two regions divided by the atlas regions' volume is used to quantify the structural connectivity.  

Functional volumes are spatially smoothed with a Gaussian kernel (5 mm full-width at half-maximum). 
The first 10 volumes are discarded so that the fMRI signal achieves steady-state magnetization, resulting in $P = 1190$ time points. Voxel fMRI time courses are detrended and band-pass filtered [0.01 - 0.15] Hz 
to improve the signal-to-noise ratio for typical resting-state fluctuations. Finally, Glasser’s multimodal parcellation (the same used for the structural connectome) resliced to fMRI resolution is used to parcellate fMRI volumes and compute regionally averaged fMRI signals. These were z-scored and stored in an $N \times P$ matrix. The functional connectivity network for each subject is also constructed by computing the pairwise Pearson correlation between the time-series data corresponding to each region. 
For baseline comparison with respect to functional connectivity based hub node detection, node strengths, i.e., degrees, of the functional connectome are computed as the sum of absolute correlation values.
%
\vspace{-0.1in}
\subsection{Hub Node Detection}
\begin{figure}
    \centering
    \begin{subfigure}[b]{0.49\linewidth}
    \includegraphics[width=0.98\linewidth]{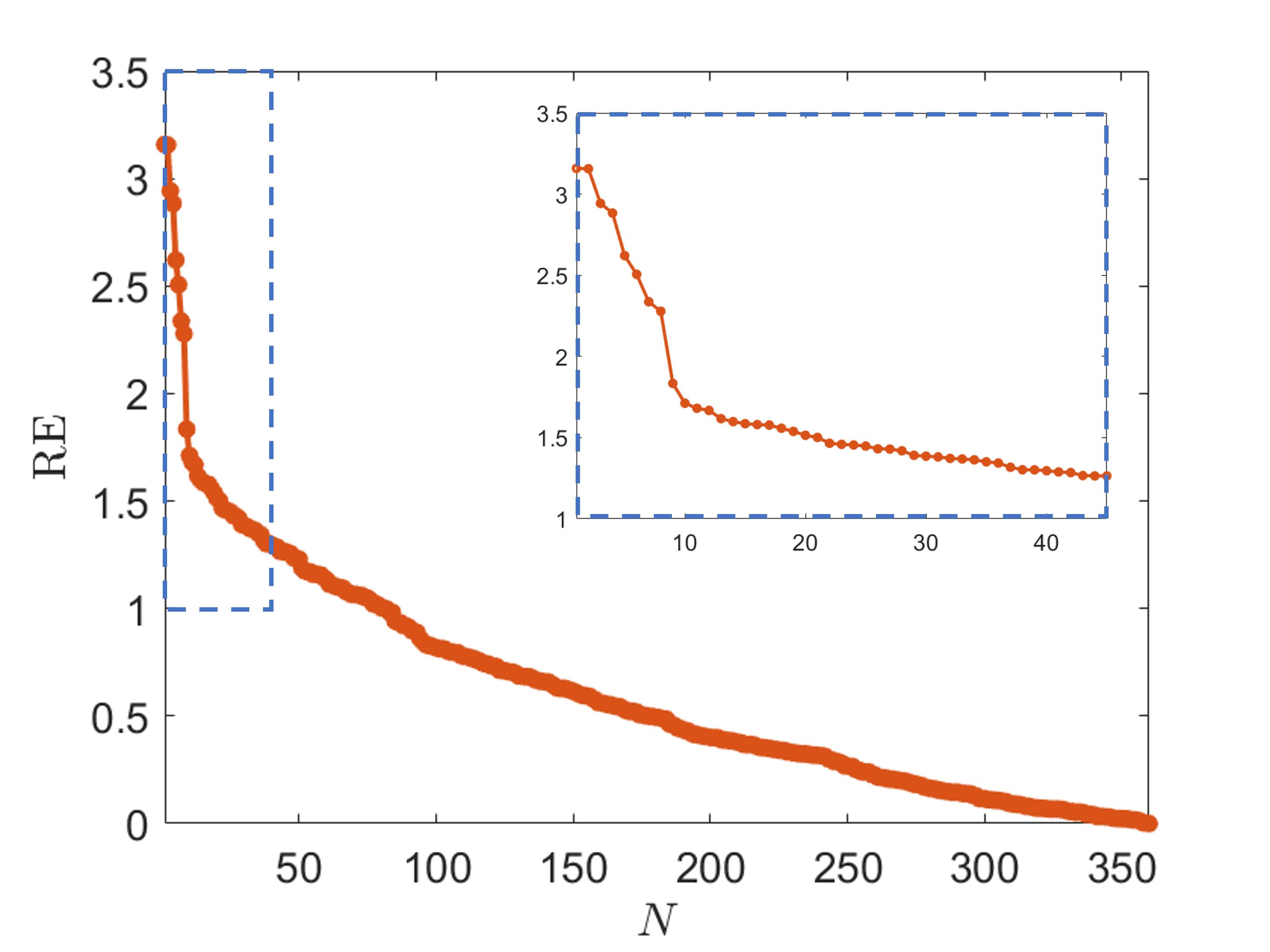}
    \caption{}
    \end{subfigure}
    \begin{subfigure}[b]{0.49\linewidth}
    \includegraphics[width=0.98\linewidth]{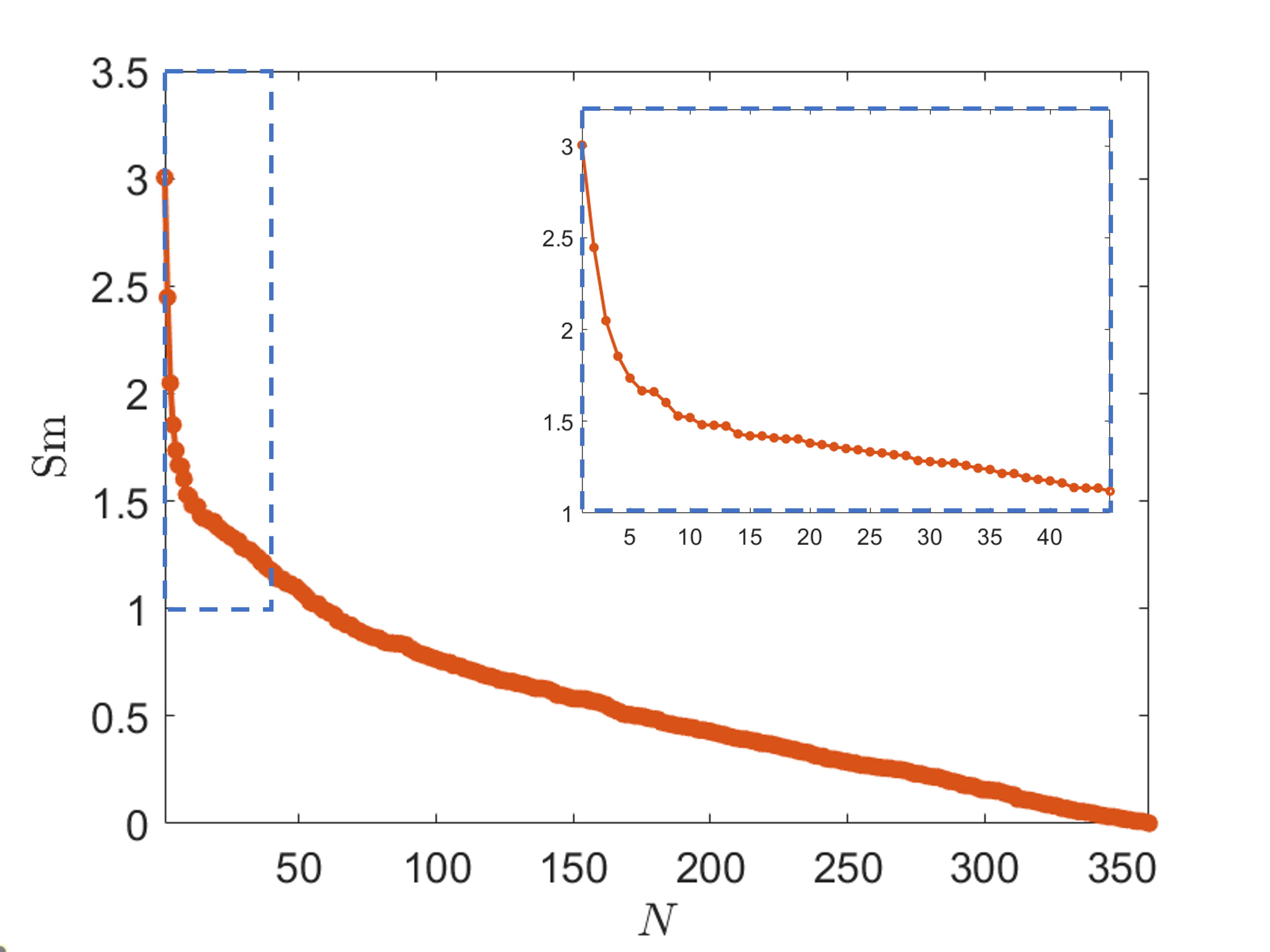}
    \caption{}
    \end{subfigure}
    \caption{Selection of $K$ for (a) Reconstruction Error; (b) Smoothness scoring. The average hub scores across subjects are sorted in decreasing order for the two methods.}
    \label{fig:kgap}
\end{figure}
GraFHub with the two anomaly scoring functions, i.e., $\textrm{GraFHub}_{\text{RE}}$ and $\textrm{GraFHub}_{\text{Sm}}$, are applied to HCP data  from two sessions, where the structural networks correspond to the graphs and the BOLD signals to the graph signals. The hub nodes for all subjects and sessions are detected separately based on thresholding or top $K$-hub node methods  as described in Section \ref{ssec:hubscore}.  For the thresholding method, hubs are nodes with z-score greater than 3. For the top $K$-hub node method, the value of $K$ is determined by calculating the average hub score of each node across subjects, and $K$ is set to the value where there is a significant drop in the average hub score, as shown in Figure \ref{fig:kgap}. Based on Figure \ref{fig:kgap}, we select $K=9$ and $K = 8$ for reconstruction error and smoothness scoring functions, respectively.

The detected hub nodes are further filtered through the participation coefficient to identify the connector hubs. For this purpose, the Louvain algorithm \cite{blondel2008fast} is applied to each subject's structural connectivity graph to detect the community structure. Each node's participation coefficient, which quantifies how evenly a node's connections are distributed with respect to community structure, is computed as \cite{guimera2005cartography}:
\begin{align}
    P_i = 1 - \sum_{s=1}^{N_M} {(\frac{k_{is}}{k_i})}^2,
\end{align}
where $N_M$ is number of identified modules (communities), $k_i$ is degree of node $i$ and $k_{is}$ is the total strength of the connections node $i$ makes with nodes in module $s$. The hub nodes with participation coefficient $0.35 < P < 0.72$ are considered as connector hubs \cite{guimera2005cartography}.

The optimal values for the filter order, $T$, and the hyperparameter, $\alpha$, in \eqref{eq:probform} are determined based on the consistency of hubs across subjects inspired by  \cite{xu2022meta}. For each $(T, \alpha)$ pair, a $56 \times 22$ matrix is constructed where each row indicates the number of hub nodes in a brain cortex (based on the Glasser parcellation) for a given subject across the two sessions. The $56 \times 56$ correlation matrix, $\matr{C}$, quantifying the pairwise correlation between subjects is then constructed and $\|\matr{C}\|_{F}$ is computed to quantify the average correlation of hub nodes across subjects. The $(T, \alpha)$ pair that yields the highest norm is selected. For the smoothness scoring function with thresholding, the optimal parameters are found as $T = 6$ and $\alpha = 5$. For reconstruction error scoring with thresholding, the optimal parameters are $T = 4$ and $\alpha = 0.5$. For the top $K$-hub method, the optimal parameters for the smoothness and reconstruction error scoring functions are $T = 6$ and $\alpha = 0.2$ and $T = 4$ and $\alpha = 2$, respectively.

\begin{figure}
    \centering
    \vspace{-0.1in}
    \includegraphics[width=1\linewidth]{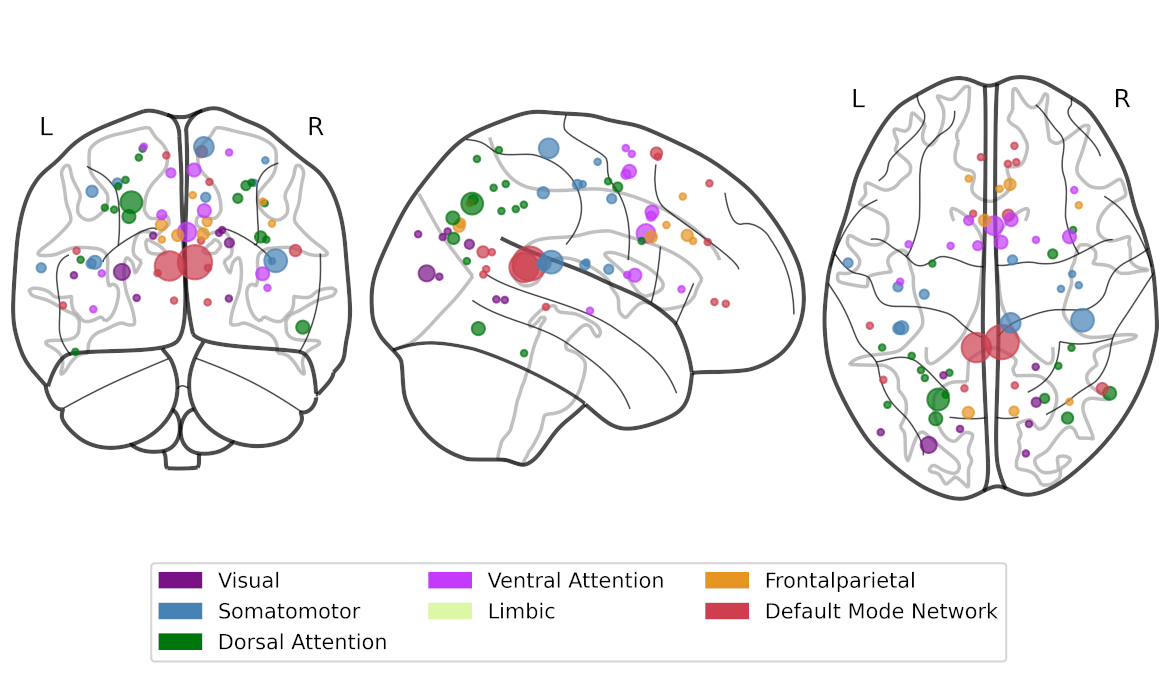}
    \caption{Consistency of the top-$K$ hubs detected by GraFHub$_{\text{Sm}}$ across all subjects plotted over the brain topomap.  The size and the color  of the nodes correspond to the number of times across 56 subjects a particular node has been detected as a hub and the brain network (Yeo's parcellation networks) to which the node belongs, respectively. }
    \label{fig:topomap}
\end{figure}

Figure \ref{fig:topomap} illustrates the top-$K$ connector hubs detected by GraFHub$_\text{Sm}$  across all subjects for one run. In particular, the size of the hub denotes how many times a particular node has been detected as a hub across 56 subjects. The color of the hub denotes the resting state brain network, determined by Yeo's parcellation \cite{yeo2011organization}, the node belongs to. From this figure, it  can be seen that nodes in the default mode network (DMN), frontal parietal and dorsal attention networks are consistently selected as hubs across subjects.

For both hub node scoring and detection methods for GraFHub, the percentage of hubs within a brain network across two sessions and all subjects are calculated similarly to a recently published meta-analysis of hub nodes in rs-fMRI \cite{xu2022meta}. While the hub detection is performed on the Glasser parcellation with $360$ regions, the percentage of hubs is reported for larger brain networks determined by Yeo's parcellation with 7 networks (Table \ref{tab:yeo}). We compare the different implementations of GraFHub with the methods discussed in Section \ref{sec:experiments}. For centrality-based methods, i.e., degree, eigenvector, closeness and betweenness, nodes with top-10 hub scores are selected as hubs. For methods that are only based on graph connectivity, i.e.,  centrality-based measures, GHFC and GFT-C, hubs are further filtered using participation coefficient to find connector hubs similar to GraFHub. For LOF and Isolation Forest, all detected hubs are used since they only employ graph signals.  

\setlength{\tabcolsep}{3pt}
\begin{table*}[h]
\centering
\resizebox{\textwidth}{!}{
\begin{tabular}{lccccccccccccccc}
\hline
Networks (area\%) & Degree & Eigenvector & Closeness & Betweenness & LOF & Isolation Forest & GHFC & GFT-C & Meta-Analysis \cite{xu2022meta} & GraFHub$_\text{RE}$ & GraFHub$_\text{RE-K}$ & GraFHub$_\text{Sm}$ & GraFHub$_\text{Sm-K}$\\
\hline\hline
Visual(14.8\%) & 38.4 & 40.6 & 14.2 & 9.2 & 15.8 & 4.5 & 4.8 & 31.40 & 9.9 & 66.7 & 18.7 & 8.0 & 6.6\\
Somatomotor(20.2\%) & 24.9 & 24.9 & 9.8 & 7.7 & 0.9 & 3.9 & 15.9 & 4.61 & 14.4 & 0.0 & 29.0 & 16.0 & 19.0\\
Dorsal Attention(11.4\%) & 12.2 & 11.6 & 8.8 & 5.5 & 3.4 & 2.8 & 15.5 & 11.61 & 16.5 & 16.7 & 6.5 & 14.9 & 17.5\\
Ventral Attention(12.1\%) & 16.0 & 16.0 & 11.3 & 8.8 & 9.4 & 9.8 & 23.2 & 5.06 & 15.6 & 0.0 & 14.0 & 13.3 & 15.2\\
Limbic(7.8\%) & 0.1 & 0.0 & 22.6 & 35.6 & 63.4 & 52.2 & 0.5 & 21.43 & 0.2 & 0.0 & 22.4 & 0.0 & 0.0\\
FrontalParietal(12.9\%) & 3.5 & 3.0 & 9.9 & 8.1 & 1.9 & 5.2 & 7.7 & 19.05 & 15.9 & 16.7 & 2.8 & 7.4 & 9.4\\
Default(20.8\%) & 5.1 & 4.0 & 23.4 & 25.1 & 5.2 & 21.6 & 32.4 & 6.85 & 27.5 & 0.0 & 6.5 & 40.4 & 32.4\\
\hline
\end{tabular}
}
\caption{Percentage of hub nodes detected in brain networks defined by Yeo's parcellation \cite{yeo2011organization}.}
\label{tab:yeo}
\end{table*}

\begin{figure*}
    \centering
    \begin{subfigure}[b]{0.24\linewidth}
        \includegraphics[width=0.98\linewidth]{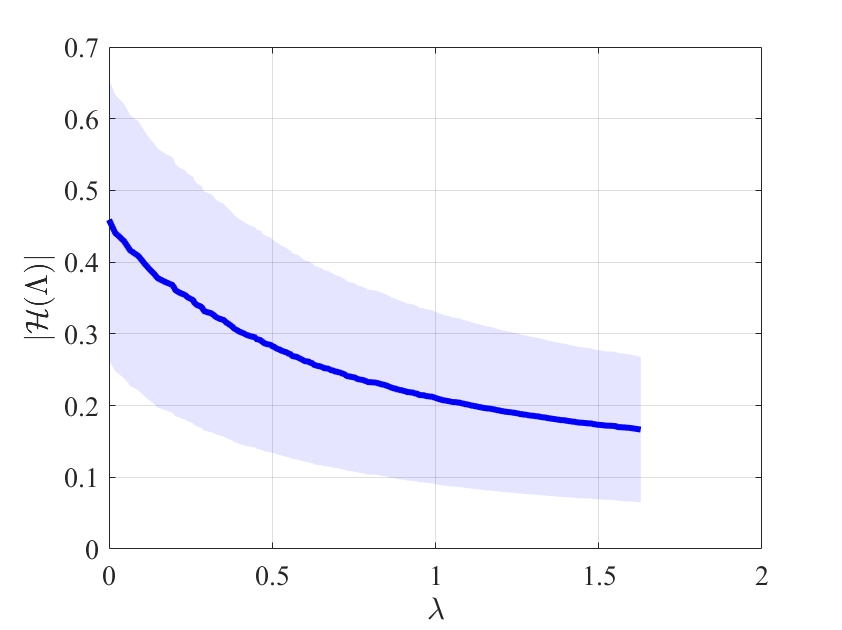}
        \caption{}
         \label{fig:freqresponse}
    \end{subfigure}
    \begin{subfigure}[b]{0.24\linewidth}
    \includegraphics[width=0.98\linewidth]{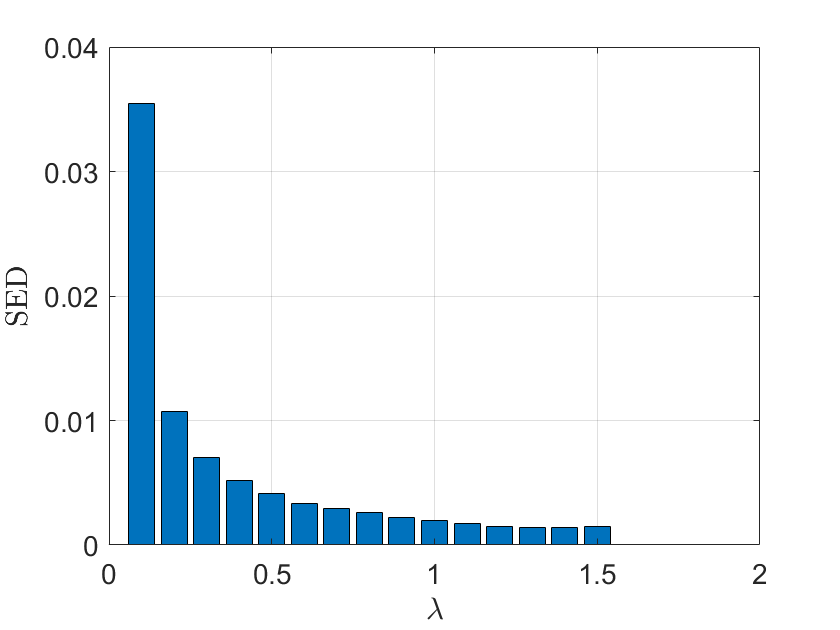}
    \caption{}
    \label{fig:sed}
    \end{subfigure}
    \begin{subfigure}[b]{0.24\linewidth}
        \includegraphics[width=0.98\linewidth]{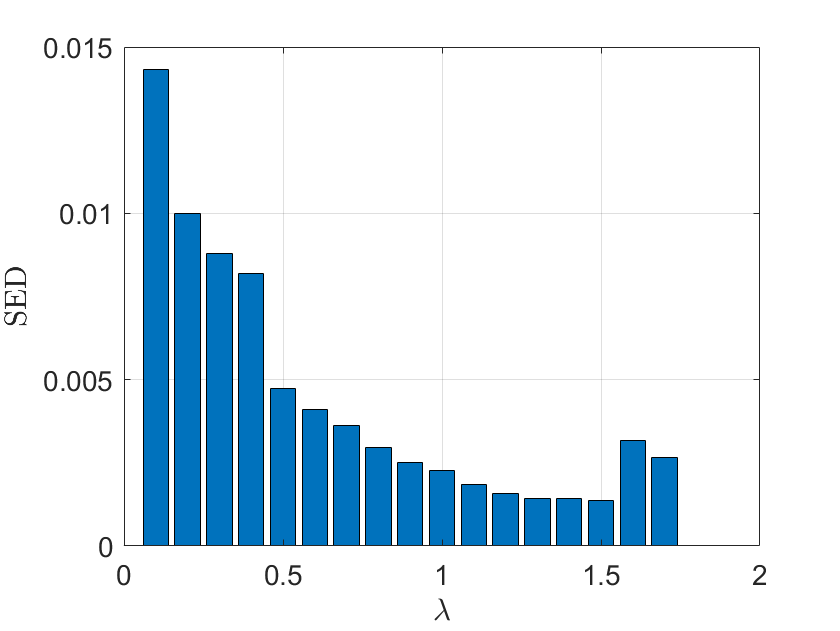}
        \caption{}
         \label{fig:subject13}
    \end{subfigure}
    \begin{subfigure}[b]{0.24\linewidth}
        \includegraphics[width=0.98\linewidth]{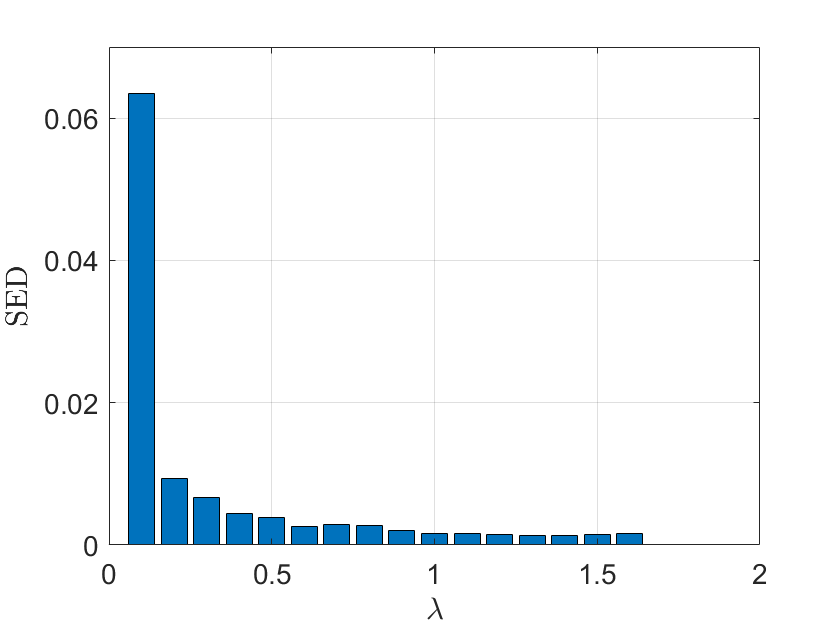}
        \caption{}
         \label{fig:subject36}
    \end{subfigure}
    \caption{(a) Average filter response across subjects. (b) Average Spectral Energy Distribution (SED) of the graph signal $\matr{F}$ across all subjects. (c) SED of the graph signal $\matr{F}$ of subject 13 (Average Hub Score = 3.3654 (RE) and 3.7354 (Sm)). (d) SED of the graph signal $\matr{F}$ of subject 36 (Average Hub Score = 2.6225 (RE) and 2.7851 (Sm)). }
\end{figure*}

\vspace{-0.1in}
\subsection{Frequency Response of the Learned Filters}

In Figure \ref{fig:freqresponse},  the frequency response of the learned filter, $\mathcal{H}(\matr{\Lambda})$, averaged across subjects, is shown. As expected, the frequency response is low-pass since $\tilde{\matr{F}}$ corresponds to the normal or smooth activity on the graph (Assumption \ref{ass:diff}). The standard deviation of the filter across subjects shows that while there's subject variation in the magnitude response of the filter, the overall shape across subjects does not change.
\vspace{-0.05in}

In order to better understand and interpret the frequency response of the learned filters, we examine the Spectral Energy Distribution (SED) of the graph signals with respect to the eigenvalues of the graph normalized Laplacian. For a graph signal $\matr{F}$ and its GFT $\widehat{\matr{F}}=\matr{U}^\top\matr{F}$, the spectral energy distribution at $i$-th eigenvalue $\lambda_i$ is defined as the average of $\widehat{F}_{ip}^2/\sum_{i=1}^N (\hat{F}_{ip})^2$ across the $P$ graph signals. Figure \ref{fig:sed} shows the average SED of the graph signals across all subjects. As it can be seen, the average of the spectral energy distribution across all subjects is mostly localized in the low-frequency range, similar to the learned filter in Figure \ref{fig:freqresponse}. This alignment of the filter shape with the SED profile is expected as the filter is learned to capture the normal activity, i.e., low-frequency content.
\vspace{-0.05in}

In order to further investigate the relationship between the underlying signal's graph spectrum and the learned hub nodes, we compute the ratio of the total SED in the high-frequency band to the total SED in the low-frequency band as $\sum_{\lambda_i\geq 1}\text{SED}(\lambda_i)/\sum_{\lambda_i< 1}\text{SED}(\lambda_i)$ for each subject. In addition, the hub activity for each subject is quantified by the average score of the top-$K$ hub nodes for both $\textrm{GraFHub}_{\text{RE}}$ and $\textrm{GraFHub}_{\text{Sm}}$. Figure \ref{fig:subject13} shows the SED of subject 13, which is the subject with the highest hub activity and the highest SED ratio for both $\textrm{GraFHub}_{\text{RE}}$ and $\textrm{GraFHub}_{\text{Sm}}$. On the other hand, Figure \ref{fig:subject36} shows the SED of subject 36, which is  the subject with the lowest hub scores and the lowest SED ratio. From these figures, it can be seen that subject 13 has significant high-frequency activity compared to subject 36, whose SED is mostly concentrated in the low frequencies. The average hub activity z-scores for subject 13 are 3.3654 and 3.7354 for $\textrm{GraFHub}_{\text{RE}}$ and $\textrm{GraFHub}_{\text{Sm}}$, respectively, compared to the average hub activity z-scores across all subjects 2.8114 ($\textrm{GraFHub}_{\text{RE}}$) and 3.0959 ($\textrm{GraFHub}_{\text{Sm}}$). On the other hand, the hub activity z-scores for subject 36 are 2.6225 and 2.7851 for $\textrm{GraFHub}_{\text{RE}}$ and $\textrm{GraFHub}_{\text{Sm}}$, respectively. These results validate Assumption \ref{ass:smooth} as the subjects with high hub activity have more high-frequency content.

\vspace{-0.1in}
\subsection{Inter-Subject Variability}
The consistency of the hubs detected by GraFHub across subjects is quantified using  normalized entropy. For each cortex $r$, we construct a vector $\matr{x}^{r}\in \mathbb{R}^{56\times 1}$ whose entries correspond to the number of hub nodes within cortex $r$ for a particular subject. After normalizing this vector, $p^{r}_{i}=\frac{x_i^{r}}{\sum_{i=1}^{56}x_i^{r}(i)}$, we calculate the normalized entropy for cortex $r$ as:
\begin{equation}
S^{r} = -\frac{\sum_{i=1}^{56} p_i^{r} \log_{2}(p_{i}^{r})}{\log_{2}(56)}.
\end{equation}
The higher the entropy, the more consistent the number of hubs is in that cortex across subjects. 

In order to quantify the significance of the normalized entropy estimates, we utilize  bias-corrected and accelerated (BCa) bootstrapping with 9,999 samples and calculate the normalized entropy of each sample. 
We report the cortices with significant normalized entropy values at the 95\% confidence interval in Table \ref{tab:glasser}. 

Cortices with high normalized entropy correspond to regions with higher consistency across subjects. For example, posterior cingulate cortex (PCC) has the highest normalized entropy for GraFHub$_{Sm}$ and is part of the DMN which has the highest percentage of hubs according to Table \ref{tab:yeo}. Similarly, anterior cingulate and medial prefrontal cortices have high normalized entropy and correspond to frontoparietal and ventral attention networks. Thus, the statistical significance of the brain networks with high percentage of hub nodes in Table \ref{tab:yeo} is established through normalized entropy as cortices with significantly high normalized entropy correspond to these networks. 

\setlength{\tabcolsep}{3pt}
\begin{table}[h]
\centering
\resizebox{0.48\textwidth}{!}{
\begin{tabular}{l*{4}{c}}
\hline
Cortices & GraFHub$_\text{RE}$ & GraFHub$_\text{RE-K}$ & GraFHub$_\text{Sm}$ & GraFHub$_\text{Sm-K}$\\
\hline\hline
Anterior Cingulate and Medial Prefrontal& $-$ & 0.26 & 0.64\text{*} & 0.69\text{*}\\
Auditory Association & $-$ & $-$ & 0.00 & 0.00\\
Dorsal Stream Visual & $-$ & 0.00 & 0.46 & 0.51\\
Dorsolateral Prefrontal & $-$ & 0.00 & 0.17 & 0.16\\
Early Auditory & $-$ & 0.39 & 0.50 & 0.56\\
Early Visual & 0.26 & 0.39 & $-$ & $-$\\
Inferior Frontal & 0.00 & 0.00 & $-$ & 0.00\\
Inferior Parietal & $-$ & 0.27 & 0.34 & 0.39\\
Insular and Frontal Opercular & $-$ & 0.41\text{*} & 0.00 & 0.26\\
Lateral Temporal & $-$ & 0.34\text{*} & 0.00 & 0.16\\
MT+ Complex and Neighboring Visual Areas & $-$ & 0.17 & 0.33 & 0.43\\
Medial Temporal & 0.00 & $-$ & $-$ & $-$\\
Orbital and Polar Frontal & $-$ & 0.57\text{*} & $-$ & $-$\\
Paracentral Lobular and Mid Cingulate & $-$ & 0.00 & 0.17 & 0.34\\
Posterior Cingulate & $-$ & 0.47 & 0.80\text{*} & 0.86\text{*}\\
Posterior Opercular & $-$ & 0.26 & 0.16 & 0.26\\
Premotor & $-$ & 0.00 & 0.00 & 0.16\\
Primary Visual & $-$ & $-$ & $-$ & $-$\\
Somatosensory and Motor & $-$ & 0.62\text{*} & 0.50 & 0.57\\
Superior Parietal & $-$ & 0.17 & 0.56\text{*} & 0.60\\
Temporo Parieto Occipital Junction & $-$ & $-$ & $-$ & $-$\\
Ventral Stream Visual & $-$ & $-$ & $-$ & $-$\\
\hline
\end{tabular}
}
\caption{Normalized entropy of the hub nodes determined by GraFHub for each cortex. $-$ refers to cortices where no hub nodes are detected for any of the subjects. $*$ refers to cortices that have statistically significant normalized entropy values.}
\label{tab:glasser}
\end{table}

%
\vspace{-0.2in}
\subsection{Verification of Hubs Through Global Efficiency}
%
Global Efficiency (GE) is a metric that characterizes the efficiency of a parallel working system, where all the nodes in the network exchange information simultaneously \cite{latora2003economic, achard2007efficiency} and is, therefore, a measure of integration and global communication efficiency. Given a graph, GE is
defined in terms of connectivity strength, and reflects the efficiency of the interaction across the whole graph \cite{rubinov2011weight, rubinov2010complex}. GE is defined as the average inverse
shortest path length in the network, which is inversely related
to the characteristic path length:
\begin{equation}
\text{GE} = \frac{1}{N(N-1)} \sum_{i,j \in V, i \neq j} \frac{1}{d_{ij}}
\end{equation}
where $d_{ij}$ is the shortest path between nodes $i$ and $j$. A small-world network will have GE
greater than  a regular lattice but GE less than
a random network. 

In order to verify that the detected hub nodes are indeed important for the overall information processing in the brain network, we calculate the change in GE when a hub node is removed from the network. In particular,  we remove each node and its connections from the network and calculate the GE of the resulting network. We then compute the difference in the global efficiency of the network before and after node removal. We repeat this procedure for each node of the network. A larger difference in GE implies that the removed node is important for information processing, i.e.,  it has ``hub-like" characteristics. 
For each subject,  we calculate the average difference in global efficiency of the network before and after removing normal and hub nodes. The average difference in GE for hub nodes and normal nodes across subjects is then calculated along with the standard deviation (see Table \ref{tab:ge}).

\begin{table}[h!]
    \centering
    \small
    \begin{tabular}{cc}
        \hline
        $\Delta$ GE remove normal node & $\Delta$ GE remove hub node \\
        \hline \hline
        $(5.88 \pm 1.31) \times 10^{-5}$ & $(7.60 \pm 4.02) \times 10^{-4}$ \\
        \hline
    \end{tabular}
    \caption{Average difference in GE for removing normal nodes compared to hub nodes detected by GraFHub.}
    \label{tab:ge}
\end{table}

As expected, the difference in global efficiency for removing hub nodes is larger than removing normal nodes for every subject. The average difference is about 12.1 times the global efficiency loss when normal nodes are removed. This result shows that the detected hub nodes are indeed more important for information processing in the brain and contribute more to the small-world characteristics. 

\vspace{-3pt}
\section{Discussion}

From Table \ref{tab:yeo}, it can be seen that graph-theoretic methods such as degree and eigenvector centrality  detect hub nodes that are concentrated in the visual and somatomotor networks. These networks comprise the primary sensory-motor cortices and have been shown to have high global connection in prior studies \cite{cole2010identifying}. This high connectivity may be either due to the relatively large size of the network or reflect the privileged placement of visual processing in the human brain \cite{ungerleider1994and}. On the other hand, closeness and betweenness centrality measures detect hub nodes in the limbic and default mode networks. While the DMN is known to be a critical network during resting state \cite{de2013connectivity}, less is known about the limbic network. Similarly, methods that only utilize the functional BOLD signals primarily detect hub nodes in the limbic network. This network consists of regions outside the cerebral cortex and is important for emotion, reward, and other valence processing \cite{padoa2006neurons}. Unfortunately, the areas that form this resting state network are usually poorly visualized with fMRI due to nearby portions of the skull creating susceptibility artifact \cite{seitzman2019state}. Thus, the BOLD signal values in this network may be very different from those in other regions, causing methods that only utilize the signals to detect these as hub nodes. 

As there is no ground truth for hub nodes, in this paper, we compare our results  to a recent harmonized meta-connectomic analysis \cite{xu2022meta} of resting-state functional MRI data of 5212 healthy young adults across 61 independent cohorts. The majority of the hub nodes detected by GrafHub$_\text{Sm}$ are in the DMN followed by somatomotor and dorsal attention networks similar to the ordering in \cite{xu2022meta}. These results are also consistent with prior studies which report components of the DMN as hubs \cite{cole2010identifying,tomasi2011association,de2013connectivity}. DMN has been noted to be active primarily in studies of resting state activity \cite{raichle2001default} and is engaged by mind wandering \cite{mason2007wandering}, prospective and retrospective self-reflection \cite{d2008self}, and memory retrieval \cite{buckner2005molecular}, suggesting that the ‘default mode’ involves ongoing processing of information for relevance to the self. In prior studies, DMN has been shown to have the highest global brain connectivity which may reflect connections necessary to implement the wide variety of cognitive functions the network is involved in. In conjunction with DMN, another large-scale network implementing a variety of cognitive functions, the cognitive control network (CCN),  is also among the highest connectivity networks \cite{cole2010identifying}. 
While our results indicate that the majority of the hub nodes are in DMN, this is followed by somatomotor, dorsal and ventral attention networks which are part of CCN.

The proposed graph filter learning framework provides additional interpretability to the importance of structure-function coupling in hub node identification. In particular, the frequency response of the optimal graph filter and the spectral energy distribution of the BOLD signal are shown to be closely related to the average hub score for a given subject. Thus, subjects whose BOLD signals have higher graph frequency content, i.e., reduced structure-function coupling, tend to have more hub node activity. 

In addition to  introducing a learning framework, the proposed approach also introduces two hub node scoring metrics and hub node detection methods. From Table \ref{tab:yeo}, it can be seen that GraFHub$_\text{RE}$ primarily detects hub nodes in the visual network. This may be due to the fact that the reconstruction error metric does not explicitly take the graph structure into account and relies on the norm of the difference between the original and filtered graph signals. On the other hand, GraFHub$_{\text{RE-K}}$ detects hub nodes in the somatomotor and visual networks which are more aligned with the meta-study and GrafHub$_\text{Sm}$. Thus, the z-score based thresholding may be eliminating some hub nodes due to the non-Gaussianity  of the reconstruction error. Comparing the thresholding and top-$K$ approaches for GrafHub$_\text{Sm}$, we can see that the top-$K$ approach  distributes  the hub nodes more evenly across resting state networks. This is due to the fact that the same number of hub nodes, $K$, is selected across subjects, treating each subject equally, while with the thresholding method, one may detect more hubs for one subject vs. another detecting only the highest activity regions, such as DMN and neglecting other important networks.  

\vspace{-2pt}
\section{Conclusions}
In this paper, we introduced a graph signal processing based framework for identifying the hub nodes in the brain. The proposed framework relies on the assumption that hub nodes are highly connected and have high activity levels with respect to their neighbors. From the perspective of GSP, this assumption results in modeling the hub nodes' activity as high-frequency with respect to the underlying graph, while the non-hub nodes have low-frequency or smooth activity. This model is implemented through an optimization problem that learns the optimal graph filter for detecting hub nodes. The proposed framework, GraFHub, is applied to both simulated and real brain network data. It is shown that GraFHub performs better than existing connectivity-based hub node identification methods for both simulated and real brain networks as it takes the coupling between the graph topology and the graph signals defined on the graph. Moreover, the learned graph filters are low-pass and the filter response is highly correlated with the spectral energy density of the signals. Thus, learning the optimal filter provides interpretability to the spectrum of the underlying graph signal and can be used as a predictor for the number of hubs in a given brain network. 

Future work will consider several extensions of the proposed framework. First, we will consider the dynamic change in hub nodes across time. It is well-known that rs-fMRI is a dynamic process, thus the hub nodes may be changing across time similar to network connectivity states \cite{allen2014tracking}. Second, we will more closely study the relationship between hub nodes and the graph frequency spectrum. For example, the contribution of different hub nodes to the SED in different frequency bands can be quantified and used as predictors for hub node identification. 
\section{Acknowledgements}
The authors would like to thank Sema Athamnah for helping with preprocessing the HCP data and constructing the functional connectivity networks.
\vspace{-3pt}
\bibliographystyle{IEEEtran}
\bibliography{refs}
\end{document}